\documentclass[reprint,showpacs,onecolumn,aps,superscriptaddress]{revtex4-2}
\usepackage{amsfonts}
\usepackage{amsmath}
\usepackage{amssymb}
\usepackage{graphicx}
\usepackage{verbatim}
\usepackage{textcomp}
\usepackage{hyperref}
\usepackage{upgreek}
\usepackage[sort&compress]{natbib}
\usepackage{threeparttable}

\providecommand{\U}[1]{\protect\rule{.1in}{.1in}}

\begin{document}
\title{Ultrasound sensing with optical microcavities}
\author{Xuening Cao}
\affiliation{Beijing National Laboratory for Condensed Matter Physics, Institute of Physics, Chinese Academy of Sciences, Beijing 100094, P. R. China}
\affiliation{University of Chinese Academy of Sciences, Beijing 100049, P. R. China}
\author{Hao Yang}
\affiliation{Beijing National Laboratory for Condensed Matter Physics, Institute of Physics, Chinese Academy of Sciences, Beijing 100094, P. R. China}
\affiliation{University of Chinese Academy of Sciences, Beijing 100049, P. R. China}
\author{Bei-Bei Li}
\email{libeibei@iphy.ac.cn}
\affiliation{Beijing National Laboratory for Condensed Matter Physics, Institute of Physics, Chinese Academy of Sciences, Beijing 100094, P. R. China}
\affiliation{Songshan Lake Materials Laboratory, Dongguan 523808, Guangdong, P. R. China}

\date{\today}

\begin{abstract}

Nowadays, ultrasound sensors are playing an irreplaceable role in the fields of biomedical imaging and industrial nondestructive inspection. Currently, piezoelectric transducers are the most widely used ultrasound sensors, but their sensitivities drop quickly when the size becomes smaller, leading to a typical sensor size at the millimeter to centimeter scale. In order to realize both high sensitivity and spatial resolution, various optical ultrasound sensors have been developed. Among them, ultrasound sensors using high-$Q$ optical microcavities have realized unprecedented sensitivities and broad bandwidth and can be mass-produced on a silicon chip. In this review, we introduce ultrasound sensors using three types of optical microcavities, including Fabry-Perot cavities, $\pi$-phase-shifted Bragg gratings, and whispering gallery mode microcavities. We introduce the sensing mechanisms using optical microcavities and discuss several key parameters for ultrasound sensors. We then review the recent work on ultrasound sensing using these three types of microcavities and their applications in practical detection scenarios, such as photoacoustic imaging, ranging, and particle detection.

\end{abstract}

\maketitle

\section{Introduction}

The versatility of ultrasound applications has been widely recognized. In industry, flow and level measurement, process control, non-destructive testing of materials, and many other parameters are measured by means of ultrasound \citep{1}. In addition, ultrasound has played an irreplaceable role in biomedical imaging. For example, ultrasound imaging is one of the most commonly used tools for early disease diagnosis due to its low cost, nonionizing radiation, and real-time capability \citep{2,3}. Ultrasound, which is not affected by the weather, is uniquely suited for applications in the field of transportation where moderate distance accuracy is required, such as reversing radar, object recognition and detection, and automatic obstacle avoidance \citep{4}. These various functions can only be achieved with suitable ultrasound sensors.

Piezoelectric transducers are the most widely used ultrasound sensors in production and clinical \citep{5} applications. The incident acoustic pressure waves deform the piezoelectric material and are measured in terms of the electric potential difference across the piezoelectric material induced by the deformation. This is the piezoelectric effect, which converts ultrasound signals into electrical signals. However, as the piezoelectric transducer achieves higher sensitivity only at resonance, the bandwidth is relatively small and it is difficult to attain high frequencies. In addition, their sensitivities drop quickly when their sizes decrease, leading to a typical sensor size at the millimeter to centimeter scale. Recent rapid advances in micromachining technology have made micro-electro-mechanical systems (MEMS) ultrasound sensors a new option. Capacitive micromachined ultrasound transducers (CMUTs) and piezoelectric micromachined ultrasound transducers allow the integration and miniaturization of ultrasound sensors with increased response bandwidth and sensitivity \citep{6}. Both of these techniques are subjected to electromagnetic interference and the opaque sensor structures make them difficult to achieve multimodal imaging.

In recent years, higher-sensitivity optical ultrasound sensors have become a new direction in the development of ultrasound sensing \citep{10,11}. Optical ultrasound sensors have experienced a continuous breakthrough in miniaturization from the beginning with free-space optical paths to optical fiber paths and now with the on-chip integration process. Optical ultrasound sensors can be classified as resonance-based or non-resonant-based according to the way they measure \citep{8}. Free space optics generally uses interference to measure ultrasound, such as Michelson interferometers \citep{9}, which use the change in optical path caused by ultrasound to measure the interferometric phase change and is a non-resonant-based measurement method. To make ultrasound sensors more portable and practical, optical fibers \citep{32,33,34} and waveguide \citep{35,FBG2} structures have been widely used. In order to further improve the ultrasound sensitivity, optical microcavities \citep{29} are often employed. Ultrasound changes their resonance frequencies by deforming the microcavities, which can be detected by monitoring the reflected or transmitted light intensity of microcavities. Benefiting from the high-$Q$ optical resonances, unprecedented phase measurement precision can be achieved, which allows ultrahigh ultrasound sensitivity. In addition, microcavities can be massively produced on a silicon chip resulting in a reduced cost, and are generally of microscale sizes allowing high spatial resolution for applications such as photoacoustic tomography. In the past few decades, various ultrasound sensing applications have been demonstrated using multiple types of optical microcavities.

In this review, we present the basics of ultrasound sensing using optical microcavities, including the sensing mechanisms as well as the key parameters of ultrasound sensors. This is necessary to understand the sensing principles and to compare the performance of different sensors. Then, we present influential and relatively new results in this field for three types of microcavities, including Fabry-Perot (F-P) cavities, $\pi$-phase-shifted Bragg gratings ($\pi$-BGs) and whispering gallery mode (WGM) microcavities. Their key parameters such as bandwidth and sensitivity are summarized, and their respective advantages and disadvantages are further compared. We also introduce their performance in practical applications. Finally, this review presents a comprehensive comparison of ultrasound sensors based on optical microcavities and further prospects for their future development.

\section{Ultrasound sensing mechanism}

An optical microcavity can be treated as a narrow-band optical filter. The resonance condition is satisfied when the optical path in the cavity is an integer multiple of the wavelength of light.
\begin{equation}
n_{\text {eff}}l =m\lambda_{\text {r}},\label{eq:01}
\end{equation}
where $n$ represents the effective refractive index and $l$ the cavity length, $m$ a positive integer number, and $\lambda_{\text {r}}$ the resonance wavelength. The optical field then resonates in the cavity, resulting in a Lorentzian-shaped resonance dip in the transmission spectrum. The linewidth of the dip depends on the optical quality ($Q$) factor of the cavity mode, which is determined by the optical losses of the cavity. A smaller optical loss gives a higher $Q$ factor and thus a narrower resonance linewidth $\delta \omega=\omega/Q_{\text {o}}$. The optical $Q$ factor can be also expressed as $Q_{\text {o}}=\omega/\kappa$, with $\omega$ and $\kappa$ being the resonance angular frequency and the optical decay rate of the mode. A higher optical $Q$ factor is desirable for sensing, as it provides a higher phase measurement precision. The depth of the dip is determined by the coupling strength of the light and the cavity. For a microcavity, an ultrasonic wave can induce either an optical resonance shift or a change in the coupling strength, both of which will induce a variation in the intracavity optical field and can be optically read out. These two sensing mechanisms through optical resonance shift and coupling strength variation are referred to as dispersive and dissipative sensing mechanisms, respectively. Besides the optical resonances, the optical microcavities also support mechanical resonances. Their response to external stimuli is further improved by a factor of mechanical quality factor $Q_{\text {m}}$, and therefore having a high mechanical $Q$ factor is also beneficial to achieving a better ultrasound sensitivity. The strong coupling between the optical mode and mechanical mode (i.e., optomechanical coupling) enables ultrasensitive sensing of multiple physical quantities \citep{2014APR,21}. The mechanical resonance enhanced sensitivity was brought to attention only very recently when high $Q$ mechanical resonators were developed in the past few years. In the following section, we use a cavity optomechanical system to interpret the physics of the dispersive and dissipative sensing mechanisms.

\begin{figure}[h!]
\begin{center}
\includegraphics[width=180mm]{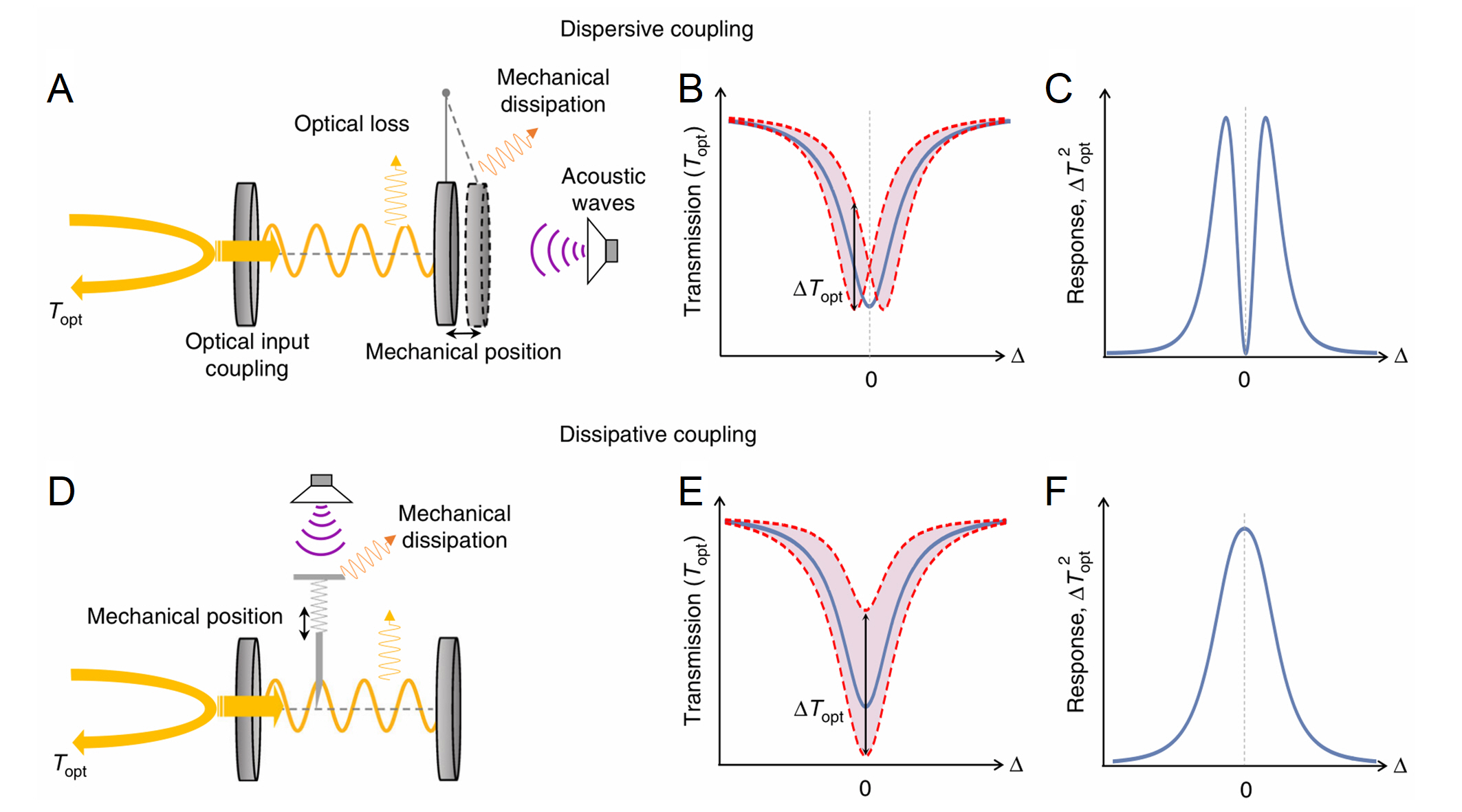}
\end{center}
\caption{Principles of dispersive and dissipative sensing mechanisms of cavity optomechanical acoustic sensing.
        \textbf{(A)(D)} Conceptual schematics of Fabry–Perot cavity-based dispersive \textbf{(A)} and dissipative \textbf{(D)} sensors.
        \textbf{(B)(E)} Cavity transmission spectrum changes in the presence of dispersive and dissipative coupling, respectively.
        \textbf{(C)(F)} Response of the dispersive and dissipative sensing mechanisms, respectively, versus the frequency (or wavelength) detuning $\Delta$ of the input laser from the cavity resonance.
        Reprint \textbf{(A-F)} from Ref. \citep{WGM15}.
        }\label{fig:1}
\end{figure}

\subsection{Dispersive sensing mechanism}

The dispersive sensing mechanism is one of the most commonly used sensing mechanisms for microcavity ultrasound sensing \citep{Dispersive1}. The principle of the dispersive sensing mechanism is illustrated in Figures \ref{fig:1}A-C. A cavity optomechanical system is shown in Figure \ref{fig:1}A, in which an optical microcavity is coupled with a mechanical resonator. When ultrasound is incident on a microcavity, the refractive index change due to the photoelastic effect and cavity length variation induced by stress will cause the resonance frequency to shift (Figure \ref{fig:1}A). This translates into a periodic modulation of the intracavity optical field at the frequency of the ultrasound. In the measurement, the frequency of the laser is usually locked on the side of the optical resonance to measure the amplitude modulation induced by the ultrasonic wave (Figure \ref{fig:1}B). For this measurement method, the optical readout response is proportional to the slope of the transmission, as shown in the response as a function of the optical frequency detuning in Figure \ref{fig:1}C. It can be seen that the response reaches the maximum when the frequency detuning $\delta \omega=\sqrt{3}\kappa/6$. The dispersive sensing mechanism can also also be read out by locking the laser frequency at the center of the optical resonance and measuring the phase modulation. An interferometer is often used to measure the phase modulation \cite{FBG6}. By modifying the two interferometric arms, this method can reduce laser phase noise. The advantage of the dispersive sensing mechanism is that the high optical $Q$ factor will amplify the optical readout signal. The higher the $Q$ factor is, the higher the readout sensitivity, and this can be applied to almost all optical microcavity sensors.

\subsection{Dissipative coupling mechanism}

Unlike the dispersive sensing mechanism that measures mode shift, the dissipative sensing mechanism relies on the change in the optical linewidth to read out the ultrasound, as shown in Figures \ref{fig:1}D-F. Ultrasound changes the total decay rate $\kappa$ by changing the rate of optical coupling into the cavity $\kappa_1$ or the intrinsic loss of the cavity $\kappa_0$. The variation in the decay rate causes variations in coupling depth (thus the output light intensity) as well as the linewidth. The optical intensity change modulated by the ultrasound can be read out by fixing the incident light frequency at the center of the resonance (Figure \ref{fig:1}E). The response reaches the maximum when the detuning $\delta \omega=0$ and decreases when the detuning increases (Figure \ref{fig:1}F). The advantage of the dissipative sensing mechanism is that some optical microcavities are not very susceptible to cavity length changes, and measuring the coupling rate changes between the cavity and the incident light can improve the response to ultrasound. For example, a recent study by Meng et~al. \citep{WGM27} found that the perimeter of the microsphere does not change significantly under ultrasound. Instead, owing to the large optical field gradient between the fiber taper and the microsphere, measuring the intensity change through the dissipative coupling can effectively improve the sensitivity.

\section{Key parameters of ultrasound sensors}

There are many parameters to evaluate the performance of ultrasound sensors, such as sensitivity, responsiveness, center frequency, bandwidth, spatial sensing capability, stability, size, etc., and different aspects will be emphasized for comparison in various application requirements. In the following, we focus on the three key parameters of ultrasound sensors, including sensitivity, working frequency and bandwidth, and spatial sensing capability, as they are more commonly used in ultrasound sensing applications. 

\subsection{Sensitivity}

Sensitivity is the most important parameter of ultrasound sensors, which is defined as the smallest detectable ultrasound pressure. It characterizes the ability to detect weak ultrasonic waves. Optical ultrasound sensors use light intensity to read out the signal, and the noise equivalent pressure (NEP) is generally used to characterize their sensitivity. NEP is the amplitude of ultrasound pressure that can be read out by the sensor at a signal-to-noise ratio (SNR) of 1. It calibrates the system noise to the effective pressure incident at the sensor surface. The bandwidth of the incident sound pressure is also relevant and NEP (with the unit of Pa) denotes the amplitude of the sound pressure with a certain bandwidth. To characterize the sensitivity of ultrasound sensors within a unit bandwidth, the noise equivalent pressure density (NEPD) can be used, with a unit of Pa Hz$^{-1/2}$, denoting the NEP for a bandwidth of 1~Hz, corresponding to a measurement time of one second. Increasing the measurement time will decrease the noise floor and therefore improve the NEP. NEP and NEPD, however, are not explicitly differentiated, and in some articles, NEPD is also named NEP.

\begin{figure}[h!]
\begin{center}
\includegraphics[width=180mm]{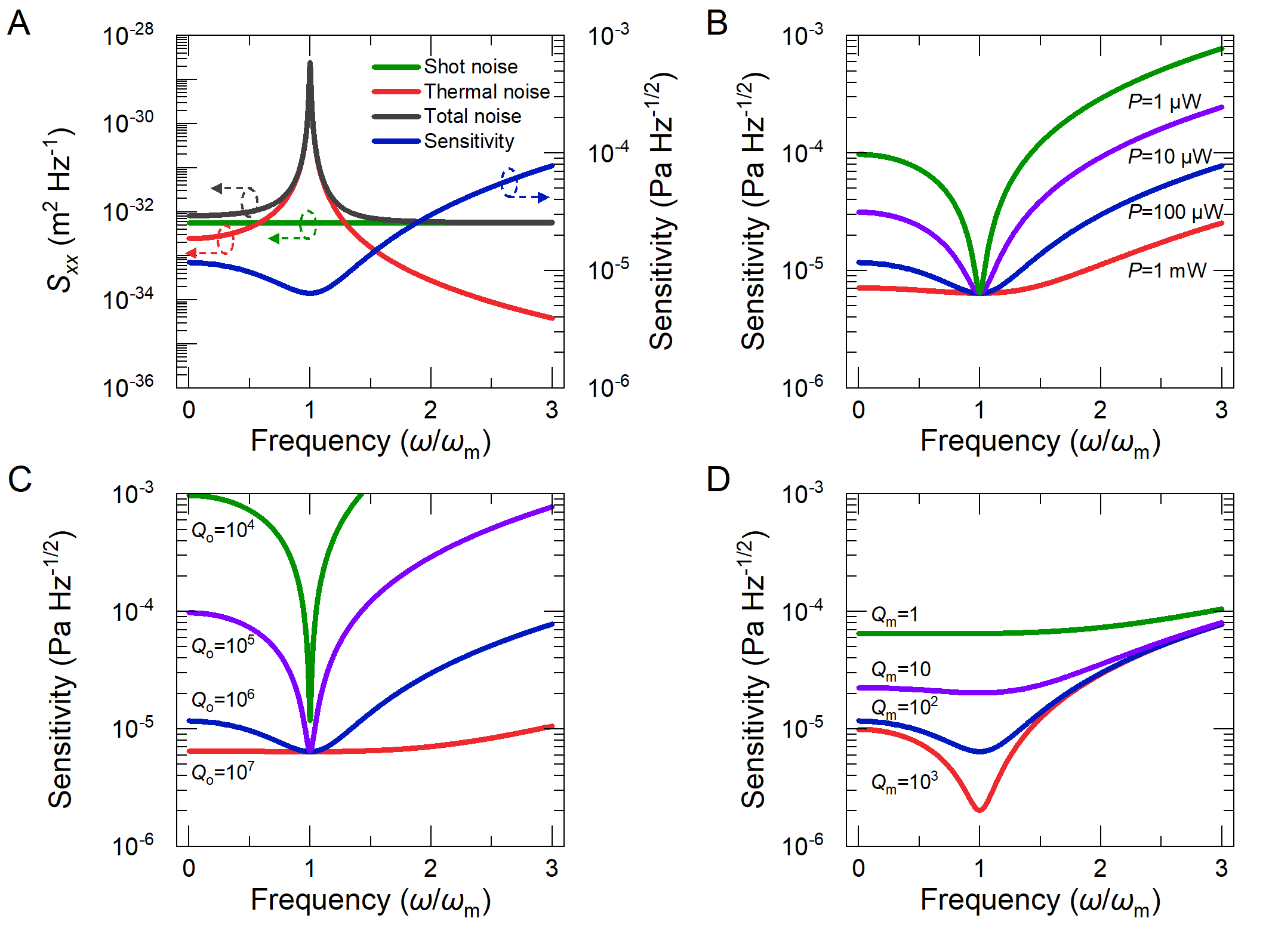}
\end{center}
\caption{
        \textbf{(A)} Displacement PSDs for the thermal noise (red curve), shot noise (green curve), total noise (black curve) on the left axis, and the corresponding sensitivity spectrum (blue curve) on the right axis. 
        \textbf{(B)} Sensitivity spectra of the microdisk ultrasound sensor, at incident optical powers of 1~\textmu W (green curve), 10~\textmu W (purple curve), 100~\textmu W (blue curve), and 1~mW (red curve), for $Q_{\text o}$ of $10^6$.
        \textbf{(C)} Sensitivity spectra for different optical quality factors $Q_{\text o}$, at an incident optical power of 100~\textmu W.
        \textbf{(D)} Sensitivity spectra at different mechanical quality factors $Q_{\text m}=1$ (green curve), $Q_{\text m}=10$ (purple curve), $Q_{\text m}=100$ (blue curve), and $Q_{\text m}=1000$ (red curve). The microdisk ultrasound sensor used here has a radius of 100~\textmu m and a thickness of 2~\textmu m. The frequency of the first-order flapping mode is 1.3~MHz.
        }\label{fig:2}
\end{figure}

As mentioned above, the ultrasound sensitivity is ultimately determined by the noise level. For a typical cavity optomechanical sensor, the main sources of noise include the thermal noise related to the environment temperature and the detection noise of the probe laser. The thermal noise mainly comes from the environmental medium damping and the intrinsic loss of the structure, and its displacement noise power spectral density (PSD) is expressed as \citep{30}
\begin{equation}
S_{x x}^{\text {thermal }}(\omega)=|\chi(\omega)|^{2} S_{FF}^{\text {thermal }}=\frac{2 \gamma k_{\mathrm{B}} T}{\left.m\left[\left(\omega_{\mathrm{m}}^{2}-\omega^{2}\right)^{2}-\omega^{2} \gamma^{2}\right)\right]}.\label{eq:04}
\end{equation}
$\chi$($\omega$) = $\frac{1}{\emph{m}(\omega_{\rm{m}}^{2}-\omega^{2}-\emph{i}\gamma\omega)}$ is the mechanical susceptibility, which quantifies the displacement of the mechanical resonator in response to an external force in the frequency domain, considering a simple case of a single mechanical resonance with a frequency of $\omega_{\text {m}}$. $m$ is the effective mass and $\gamma$ is the damping rate of the mechanical resonator. Decreasing $\gamma$ (thus increasing mechanical quality factor $Q_{\text {m}}$) can enhance the response to near-resonant forces. The detection noise includes the classical technical noise (phase noise and intensity noise) and the quantum shot noise. The technical noise can be significantly suppressed using homodyne or heterodyne detection schemes. As a result, we only consider the shot noise in the following. \citep{21} To better visualize the noise and sensitivity as a function of the frequency, a microdisk optomechanical sensor is utilized as an example. A microdisk with a radius of 100~\textmu m and a thickness of 2~\textmu m has been simulated with a mechanical resonance frequency of 1.3~MHz. The red curve in Figure \ref{fig:2}A shows the thermal noise PSD near the mechanical resonance frequency. The temperature $T$  is 300~K and the $Q_{\mathrm{m}}$ is 100. It can be seen that there is a thermal noise peak near the mechanical mode due to the resonance enhancement, and the response at the mechanical resonance is enhanced by a factor of 100. The displacement PSD of the shot noise is expressed as \citep{31}
\begin{equation}
 S_{x x}^{\text {shot }}(\omega)=\frac{\kappa}{16 \eta N G^{2}}\left(1+4\frac{\omega^{2}}{\kappa^{2}}\right). \label{eq:05}
\end{equation}
In this equation, $N=Q_{\mathrm{o}}P/\hbar \omega_{\mathrm{L}}^2$ is the intracavity photon number, where $P$ is the incident optical power and $\omega_{\mathrm{L}}$ is optical resonance frequency. $\kappa=\omega_{\mathrm{L}}/Q_{\mathrm{o}}$ is the optical power decay rate, and $\eta$ stands for the optical detection efficiency. $G=\frac{d\omega}{dx}$ is the optomechanical coupling coefficient, which quantifies how much the optical resonance frequency shift for a mechanical displacement $x$. The shot noise PSD is shown as the green curve in Figure \ref{fig:2}A with the optical power $P=100$~\textmu W and optical $Q$ factor $Q_{\mathrm{o}}=10^6$, which is a constant in the frequency range. The shot noise only increases significantly when the frequency is comparable to $\kappa$. The total noise from the sum of thermal noise and shot noise is shown in the black curve in Figure \ref{fig:2}A, indicating that the noise is dominated by thermal noise near mechanical resonance frequency and by shot noise when it is far from the mechanical resonance.

By definition, the sensitivity (or NEP) can be obtained from the noise PSD, which is expressed as
\begin{equation}
\mathrm{NEP}=\frac{1}{ \alpha A} \sqrt{\frac{S_{xx}^{\text {shot }}}{\left|\chi \right|^{2}}+S_{FF}^{\text {thermal }}}
 = \frac{1}{A\alpha}\sqrt{\frac{\kappa}{16 \eta N G^2 {|\chi|}^2}[1+4(\frac{\omega}{\kappa})^2]+2m\gamma k_B T}, \label{eq:06}
\end{equation}
where $\alpha$ is a coefficient related to the spatial overlap between the ultrasonic wave and the mechanical mode profile of the sensor, and the pressure difference between the upper and lower surfaces of the sensor. $A$ is the sensor area. The sensitivity versus frequency is shown in the blue curve in Figure \ref{fig:2}A. It can be seen that the sensitivity reaches a minimum at mechanical resonance frequency where the noise is dominated by thermal noise and is degraded in the shot-noise-limited regime. This is due to the fact that mechanical resonance not only enhances thermal noise but also improves response. However, the shot noise is not amplified. Therefore mechanical resonance helps increase the SNR. The fundamental limit of ultrasound sensors is thermal-noise-limited sensitivity. As a result, reaching the thermal-noise-limited sensitivity is essential to achieving high sensitivity for ultrasound sensors.

The thermal noise dominant regime can be reached by optimizing the parameters to reduce the shot noise or increase the thermal noise. From Eq. \ref{eq:06} we can see that increasing the probe power $P$, optical quality factor $Q_{\mathrm{o}}$ or the optomechanical coupling coefficient $G$, can reduce the contribution of shot noise. Figures \ref{fig:2}B,C show the sensitivity spectra for various incident powers when the $Q_{\mathrm{o}}$ is fixed at $10^6$ and for different $Q_{\mathrm{o}}$ when the incident power is fixed at 100~\textmu W, respectively. Both incident power and $Q_{\mathrm{o}}$ have no effect on the thermal noise term, so the minimum NEP (sensitivity at the mechanical resonance frequency) that can be achieved by the system is constant regardless of how these two parameters are varied. Both Figures \ref{fig:2}B,C show that as the incident power and $Q_{\mathrm{o}}$ increase, the shot noise decreases, and the frequency range of the thermal noise dominant regime increases, extending the detection bandwidth. It is shown that the sensitivity exhibits a flat spectrum in a frequency range at the order of $\omega_m$, for a proper choice of the incident power and the $Q_{\text {o}}$. In addition, as $S_{x x}^{\text {shot }}\propto \frac{1}{Q_{\mathrm{o}}^2}$ and $S_{x x}^{\text {shot }}\propto \frac{1}{P}$, increasing the $Q_{\text {o}}$ leads to a more effective reduction of the shot noise than increasing the incident power. 

Another way to achieve the thermal-noise-dominant regime is to improve the mechanical quality factor $Q_{\mathrm{m}}$. In the one hand, the increase of $Q_{\mathrm{m}}$ can not only make it easier for the system to reach the thermal noise dominant regime, but also improve the thermal-noise-limited sensitivity, as shown in Figure \ref{fig:2}D. On the other hand, the increase of the $Q_{\text {m}}$ will also narrow the linewidth of the mechanical peak, and the thermal-noise-dominated frequency range (i.e., the bandwidth). Due to the high $Q_{\text {o}}=10^6$, the thermal-noise-dominated regime can still be reached even when the $Q_{\text {m}}=1$.
Under this premise, a microcavity with lower $Q_{\mathrm{m}}$ can realize broadband detection but with compromised sensitivity, and a microcavity with higher $Q_{\mathrm{m}}$ can achieve better sensitivity but limited bandwidth in the region dominated by thermal noise. These results show that both optical resonance and mechanical resonance can improve the sensitivity from different aspects. The dual resonance in the cavity optomechanical system enables extremely high sensitivity and has been widely used to measure a broad variety of physical quantities \citep{21}. 

\subsection{Working frequency and bandwidth}

All acoustic waves above 20~kHz fall under the category of ultrasound, which has a broad frequency range. Ultrasound with different frequencies has unique applications, so the working frequency and bandwidth of ultrasound sensors are of great interest. For ultrasound imaging, ultrasound of higher frequencies (shorter wavelengths) is desirable as it can provide higher spatial resolution. In order to attain images with micrometer resolution, an ultrasound detector needs to have a center frequency and bandwidth in the MHz range \citep{15}. However, the simple pursuit of high frequencies is not desirable, as sound waves can also be scattered and absorbed in the medium. The absorption loss is proportional to the frequency, and the scattering loss is proportional to the frequency squared. The loss of ultrasonic waves in the air is dominated by the scattering loss, with the attenuation of a 1~MHz frequency ultrasound in the air at the order of 160~dB m$^{-1}$ \citep{16}. Therefore, balancing between penetration depth and image resolution is necessary. For other applications, the detection bandwidth often determines the performance of the ultrasound sensor. For example, in thermoacoustic and photoacoustic reconstruction, the axial resolution is inversely proportional to the detection bandwidth \citep{17,24}. An ultrasound sensor of wider bandwidth can detect more details in three dimensions. For another example, most ultrasonic ranging applications employ the time-of-flight (TOF) method, which determines position by reflecting sound waves from the surface of various objects. A larger bandwidth will result in a narrower pulse width in the time domain and therefore better precision. For some specific applications such as sonar and underwater communications, acoustic sensors at kHz frequency are required, to decrease the acoustic loss and therefore expand the detection and communication ranges.

The bandwidths of the traditional piezoelectric transducers are generally at megahertz level, with typical center frequencies at 1-100~MHz and fractional bandwidths (the ratio between the $-3~$dB or $-6~$dB bandwidth and the center frequency) of 60–80\%. The fractional bandwidth can be increased to over 100\% by using capacitive or piezoelectric micromachined ultrasound transducers, but with compromised center frequency generally at a few megahertz levels \citep{18}. For ultrasound sensors with optical microcavities, the bandwidth can be broadened to the level of mechanical resonance frequencies when it is in the thermal-noise-dominant regimes, as discussed above. A bandwidth of up to several hundred megahertz can be achieved using optical ultrasound sensors \citep{WGM6}. For optical resonance-based ultrasound sensors, the intracavity photon lifetime is one of the factors limiting the bandwidth. The lower the optical $Q$ factor is, the shorter the intracavity photon lifetime is, and the broader the bandwidth is. As a result, there is a tradeoff between sensitivity and bandwidth, regarding the choice of optical $Q$ factor. In addition, the bandwidth of the sensor is often limited by many other factors, such as mechanical construction, etc. Having a higher mechanical resonance frequency is helpful for achieving a broader bandwidth.


\subsection{Spatial sensing capability}

The spatial sensing capability of ultrasound sensors includes the ability to respond to ultrasonic waves from different directions (i.e., the acceptance angle), and also the ability to detect ultrasonic waves at different distances. Both of these are significantly influenced by the shape and size of the sensor. Usually, the sensor is most sensitive to the axial ultrasound, and the sensitivity gradually decreases as the angle of incidence increases. Piezoelectric elements are generally directional, typically offering acceptance angles below $\pm$ 20$^{\circ}$ \citep{19}, and may require acoustic lenses to increase their acceptance angles. For imaging applications, a wide acceptance angle facilitates more realistic spatial information, in which optical ultrasound sensors are more advantageous. There are various types of optical ultrasound sensors, some of which can achieve an almost full-angle response \citep{FP6}. In order to minimize ultrasound propagation loss, sensors are frequently placed close to the acoustic sources, but this near-field detection has its drawbacks. Spherical sensors that can be considered as points have a larger acceptance angle but a limited detection distance. While disk or membrane sensors are better suited for long-range detection, their angular response to high-frequency ultrasound is undesirable \citep{7}. Placing the microring detector at the acoustic far field provides a longer working distance and a broader acceptance angle, and the detection at the acoustic near field offers improved sensitivity and broader bandwidth but at the cost of a reduction in the acceptance angle \citep{20}.

\section{Optical microcavity ultrasound sensors}

There are many types of optical microcavities. Depending on the fabrication platform, they can be categorized as free space optical resonant cavities, optical microcavities on fibers, and optical microcavities on a chip. Based on the working principle, they can be classified as F-P cavities, $\pi$-BGs, WGM microcavities, and photonic crystal microcavities, etc. Some microcavities such as photonic crystal cavities are complex to manufacture and have a relatively low response to ultrasound, therefore are not suitable for ultrasound sensing. This review outlines three types of microcavities that are widely used in ultrasound sensing, namely F-P cavities, $\pi$-BGs, and WGM microcavities.

\subsection{Fabry-Perot cavities}

\begin{figure}[h!]
\begin{center}
\includegraphics[width=180mm]{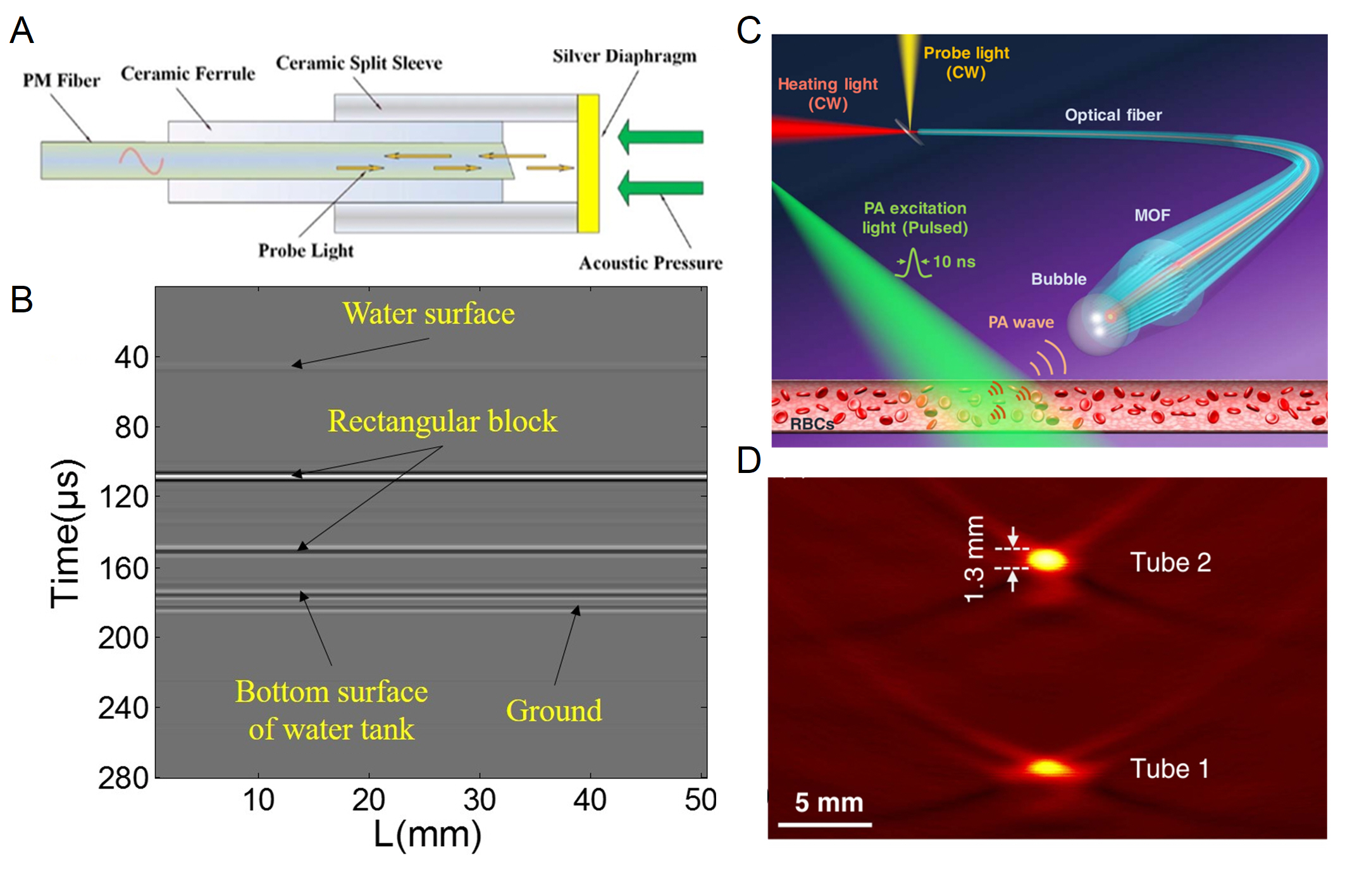}
\end{center}
\caption{Different types of F-P cavities for ultrasound sensing.
        \textbf{(A)} Schematic of the sensing head based on a large area silver diaphragm.
        \textbf{(B)} A rectangular Plexiglas block ultrasound image reconstructed by TOF approach.
        \textbf{(C)} Schematic of a surface microbubble photothermally generated at a microstructured optical fiber tip for photoacoustic imaging of red blood cells in a blood vessel.
        \textbf{(D)} Reconstructed cross-section image of the blood-filled tubes.
        Reprint \textbf{(A)} from Ref. \citep{FP3}; 
        \textbf{(B)} from Ref. \citep{FP7}; 
        \textbf{(C-D)} from Ref. \citep{FP13}. 
        }\label{fig:3}
\end{figure}

\begin{figure}[h!]
\begin{center}
\includegraphics[width=180mm]{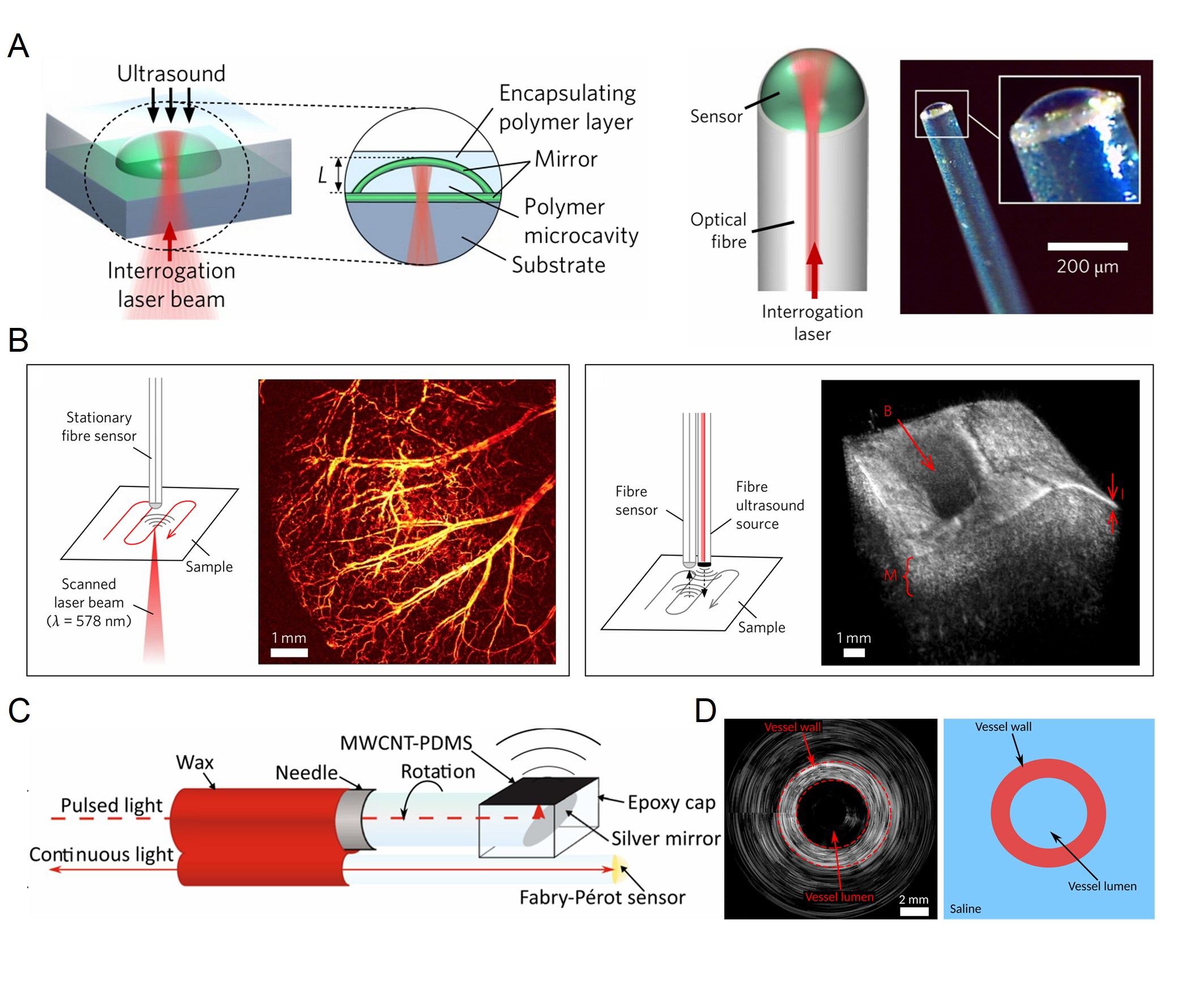}
\end{center}
\caption{F-P cavity ultrasound sensors for photoacoustic imaging.
        \textbf{(A)} Plano-concave optical microresonator ultrasound sensor (left) and the sensor integrated at the end of the optical fiber (right).
        \textbf{(B)} Schematic of fiber-microresonator-sensor based optical-resolution photoacoustic microscopy experiment and image of mouse ear vasculatures in vivo (left). On the right is shows a schematic of the all-fiber pulse-echo ultrasound experiment and 3D pulse-echo ultrasound image of ex vivo porcine aorta.
        \textbf{(C)} Schematic of the side-view optical ultrasound transducer.
        \textbf{(D)} Rotational optical ultrasound images of an ex vivo swine carotid artery (left) and corresponding schematic of the imaged vessel section (right).
        Reprint \textbf{(A-B)} from Ref. \citep{FP6};
        \textbf{(C-D)} from Ref. \citep{FP12}. 
        }\label{fig:4}
\end{figure}

\begin{figure}[h!]
\begin{center}
\includegraphics[width=180mm]{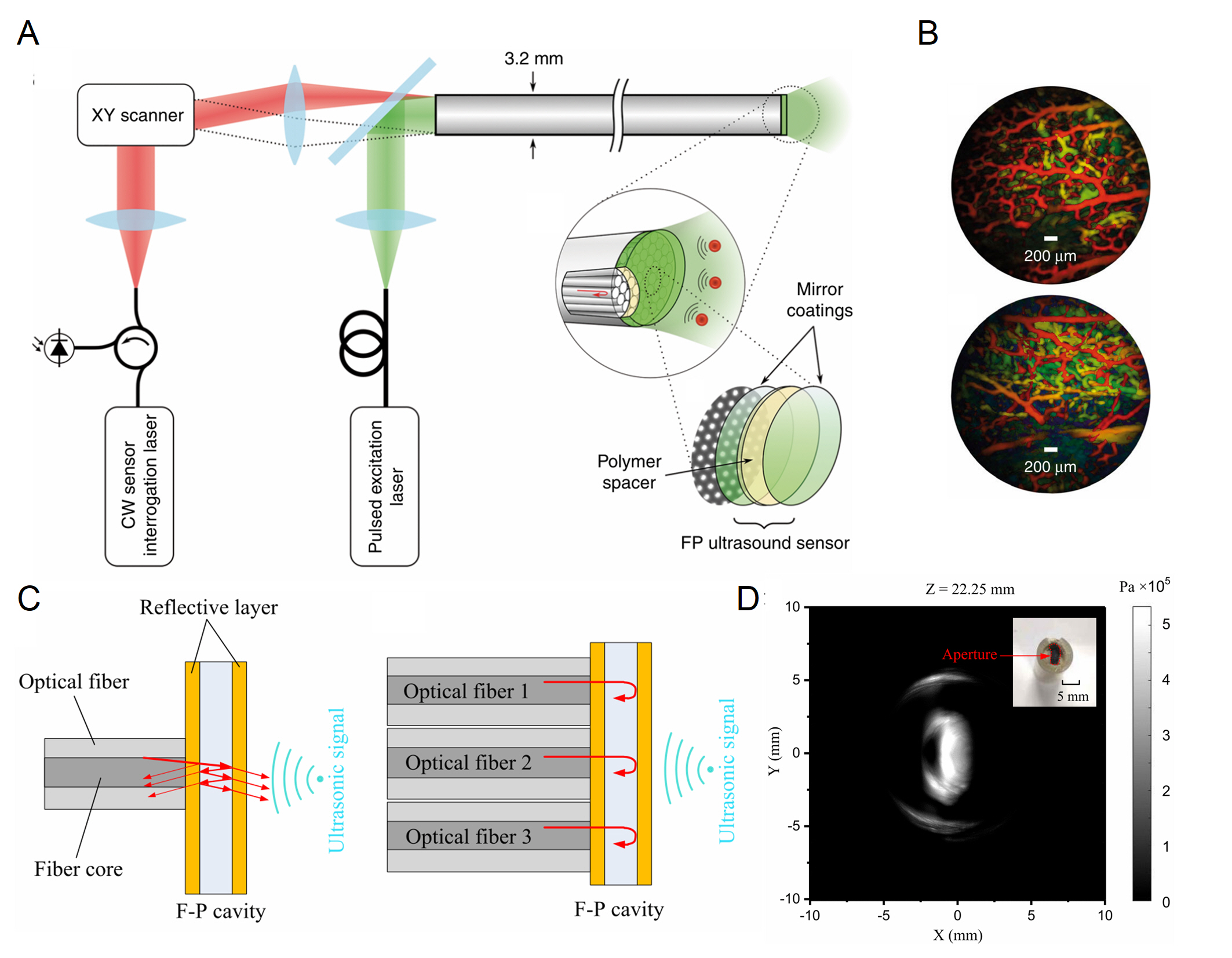}
\end{center}
\caption{Ultrasound sensing usign an array of F-P cavities.
        \textbf{(A)} All-optical forward-viewing photoacoustic endoscopy probe.
        \textbf{(B)} Photoacoustic images of an ex vivo duck embryo.
        \textbf{(C)} Schematic of F-P interferometer (left) and array (right) for ultrasound sensing.
        \textbf{(D)} Ultrasound imaging results of an arbitrary-shape ultrasound transducer.
        Reprint \textbf{(A-B)} from Ref. \citep{FP8}; 
        \textbf{(C-D)} from Ref. \citep{FP14}.
        }\label{fig:5}
\end{figure}

\begin{figure}[h!]
\begin{center}
\includegraphics[width=180mm]{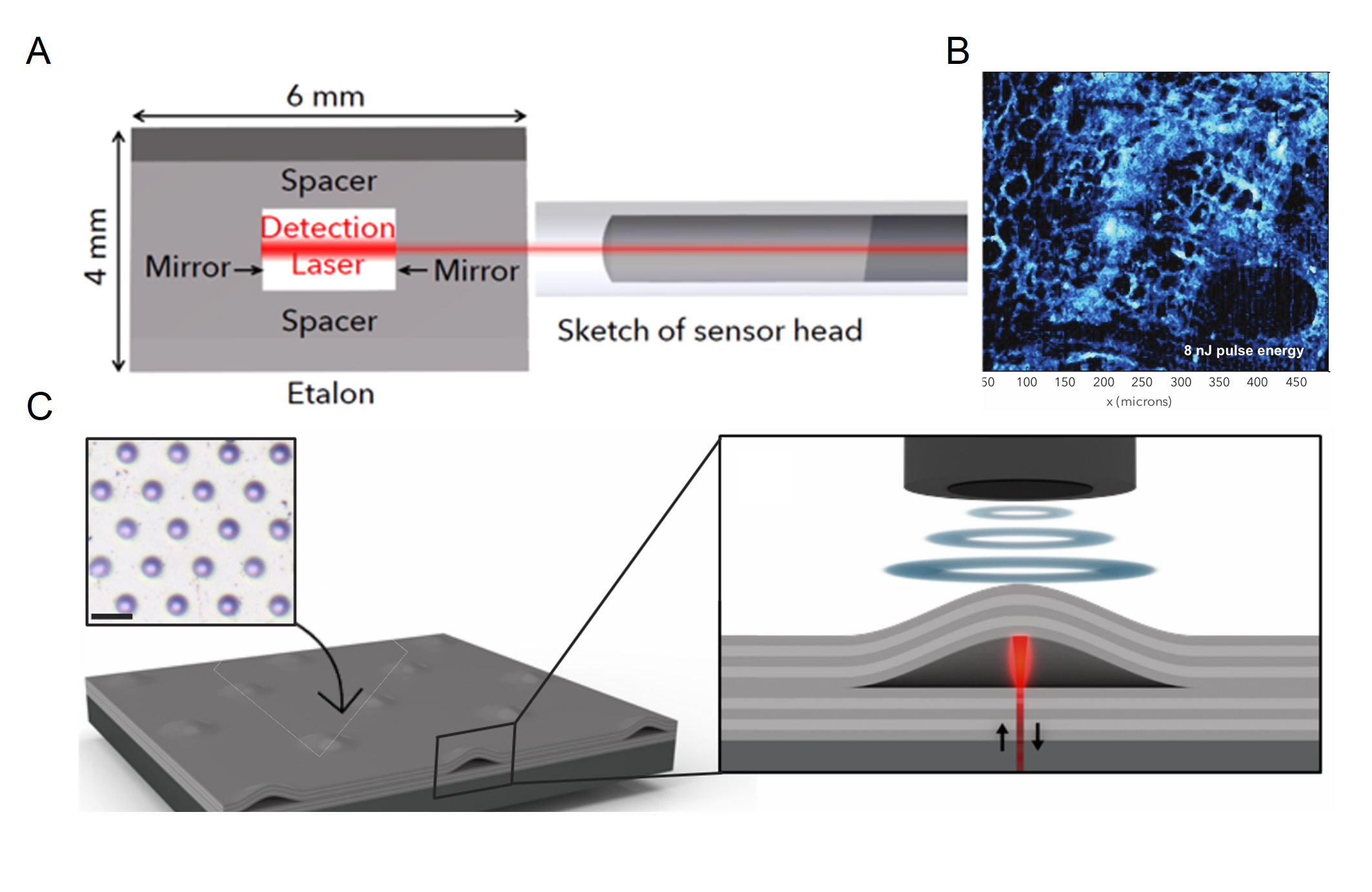}
\end{center}
\caption{
        \textbf{(A)} A miniaturized all-optical akinetic detector based on a rigid F-P resonator.
        \textbf{(B)} Photoacoustic images of Feulgen-stained Allium Cepa histology samples.
        \textbf{(C)} A depiction of an array of buckled dome etalons (left). Inset: Microscope image of an array of buckled dome F-P microcavities. On the right shows the schematic illustration of the buckled-dome ultrasound sensor.
        Reprint \textbf{(A-B)} from Ref. \citep{FP4}; 
        \textbf{(C)} from Ref. \citep{FP16}.
        }\label{fig:6}
\end{figure}

The F-P cavities are the most basic type of optical resonators and have been widely used in many ultrasound sensors \cite{FPposition,FPforce1,FPforce2,FPaccelerometer,FPphotoacoustic}. The F-P cavity uses two highly reflective mirrors to localize light in a cavity between the mirrors, which can be realized with the light propagating in the free space or in optical fibers as well as on a chip. F-P cavities for ultrasound sensing are generally not implemented using free-space light, as free-space light paths are generally less mobile and not suitable for most sensing scenarios. Therefore, a majority of ultrasound sensors based on F-P cavities are fabricated at the end of the optical fiber, replacing one mirror with a highly reflective film, which improves both the optical $Q$ factor and the response to ultrasound \citep{FP11}. The ultrasound incident on the film causes a change in cavity length and thus modulates the light intensity of the reflected light. In 2013, an F-P cavity using a multilayer graphene film as a reflector was used for ultrasound sensing \citep{FP2}. Because the film was only 100~nm thick, it realized a NEP of down to 60~\textmu Pa Hz$^{-1/2}$ at 10~kHz and a flat response in the 0.2-22~kHz frequency range. The NEP was then reduced to 14.5~\textmu Pa Hz$^{-1/2}$ by Xu et~al. using a silver film with higher reflectivity \citep{FP3}. Figure \ref{fig:3}A shows a schematic diagram of the F-P cavity ultrasound sensor with a silver film. Polymer films with smaller Young's modulus have also been used to increase the response to ultrasound. Ultrasound sensors made from 353ND \citep{FP5} and polytetrafluoroethylene (PTFE) \citep{FP7} films have enabled the TOF method ranging with resolutions of 5~mm and 3.7~mm respectively. Figure \ref{fig:3}B shows an ultrasound reconstruction of a Plexiglas block in water using a PTFE diaphragm F-P cavity. A microbubble has also been used for ultrasound sensing as shown in Figure \ref{fig:3}C \citep{FP13}. The microbubble was photothermally generated on a microstructured optical fiber tip, forming a flexible F-P cavity whose gas-water interface was sensitive to ultrasonic waves. This microbubble was capable of detecting weak ultrasounds emitted from red blood cells irradiated by pulsed laser light. Figure \ref{fig:3}D shows the reconstructed cross-section photoacoustic image of the blood-filled tubes using this microbubble cavity. To improve the chemical stability of the film and siplify the fabrication process, Fan et~al. have made an F-P cavity by splicing three sections of cleaved standard single-mode fibers with the off-core cross-section in the middle \citep{FP10}. This multi-mode dual-cavity F-P interferometer ultrasound sensor has achieved a broadband ultrasound response from 5~kHz to 45.4~MHz.

The cavity medium in the previous F-P cavities are air, which is not conducive to encapsulation and is less robust. Therefore in 2017, Guggenheim et~al. proposed a flat-concave polymer microresonator that consisted of a solid plano-concave polymer microcavity formed between two highly reflective mirrors as shown in Figure \ref{fig:4}A \citep{FP6}. With a high optical $Q$ factor of \textgreater 10$^5$, it has realized a broadband response of 40~MHz and a NEP of 1.6~mPa Hz$^{-1/2}$. This sensor can also be used as a probe for a range of applications, with an almost full angular response when combined on the end face of a fiber (Figure \ref{fig:4}A). Figure \ref{fig:4}B shows an optical-resolution photoacoustic microscopy image of the mouse ear acquired in vivo (left) and a 3D high-resolution pulse-echo ultrasound image of an ex vivo porcine aorta sample (right) using this ultrasound sensor on a fiber. Another great advantage of the sensor on the fiber is that it can penetrate deep into the tissue for endoscopic imaging. Colchester et~al. demonstrated all-optical rotational B-mode pulse-echo ultrasound imaging using an optical head at the distal end with a multi-walled carbon nanotube and polydimethylsiloxane composite coating \citep{FP12}, with its exact structure shown in Figure \ref{fig:4}C. The light pulse induces the photoacoustic effect of the coating to produce axial ultrasonic waves, and the F-P cavity next to it is used to receive the tissue echoes, thus enabling compact and minimally invasive probing. Figure \ref{fig:4}D shows the rotational optical ultrasound images of an ex vivo swine carotid artery.

Due to the special cladding-core structure of the fibers, a sensing array can be easily realized using F-P cavities \citep{FP9}. In 2018, Ansari et~al. has realized a forward-viewing endoscopic probe using a 3.2~mm diameter fiber bundle that has 50,000 cores, as shown in Figure \ref{fig:5}A \citep{FP8}. A 15~\textmu m-thick Parylene C film layer between two 90\% reflective dielectric mirrors was deposited on the end face of the fiber to form the F-P cavity. Photoacoustic tomography imaging has been realized using this probe. The excitation laser of the photoacoustic signal enters the fiber bundle from all channels, providing a large illuminated field of view. At the same time, the interrogation laser beam is scanned using a lens and coupled into different fiber cores to read out the ultrasound signals at different locations. The lateral resolution of photoacoustic imaging was related to the distance between the cores. The on-axis lateral resolution of the probe was depth-dependent and ranged from 45-170~\textmu m for depths between 1 and 7~mm, and the vertical resolution was 31~\textmu m over the same depth range. Figure \ref{fig:5}B shows the photoacoustic images of an ex vivo duck embryo in different regions. However, the drawback of multiple channels is that the optical readout is complicated because the F-P cavities in different channels not necessarily have the same resonant wavelength. In a study by Yang et~al., a photothermal tunable fiber optic ultrasound sensor array was demonstrated \citep{FP15}. The resonant wavelength of each cavity can be controlled by a laser. In 2022, Ma et~al. proposed a 4$\times$16 fiber-optic array based on F-P cavities, which enabled parallel sensing for imaging with a volume rate of 10~Hz \citep{FP14}. Furthermore, its imaging performance was characterized by reconstructing images from the multichannel signals without mechanical scanning. Figures \ref{fig:5}C,D show the principle of parallel sensing and the ultrasound imaging results of an arbitrary-shape ultrasound transducer.

In 2016, Preisser et~al. demonstrated a novel all-optical akinetic ultrasound sensor, consisting of a rigid fiber-coupled F-P etalon with a transparent central opening \citep{FP4}. Unlike other ultrasound sensors based on F-P cavities that use the displacement of the cavity mirror for ultrasound sensing, it uses the change in refractive index within the fluid-filled cavity. The sensor not only achieved a broadband resonance-free flat response in the 22.5~MHz range but also achieved a sensitivity of 450~\textmu Pa Hz$^{-1/2}$ and was used in photoacoustic imaging of biological samples as shown in Figures \ref{fig:6}A,B. In addition to being integrated on optical fibers, F-P cavities can also be integrated on a chip. Recently, Hornig et~al. realized ultrasound sensing using a monolithic buckled-dome cavity (Figure \ref{fig:6}C), with a NEP as low as 30-100~\textmu Pa Hz$^{-1/2}$ in the frequency range below 5~MHz \citep{FP16}. Due to the sensitive response of the buckled film to external forces, this device has achieved thermal-noise-limited sensitivity. 

\subsection{$\pi$-phase-shifted Bragg gratings}

\begin{figure}[h!]
\begin{center}
\includegraphics[width=130mm]{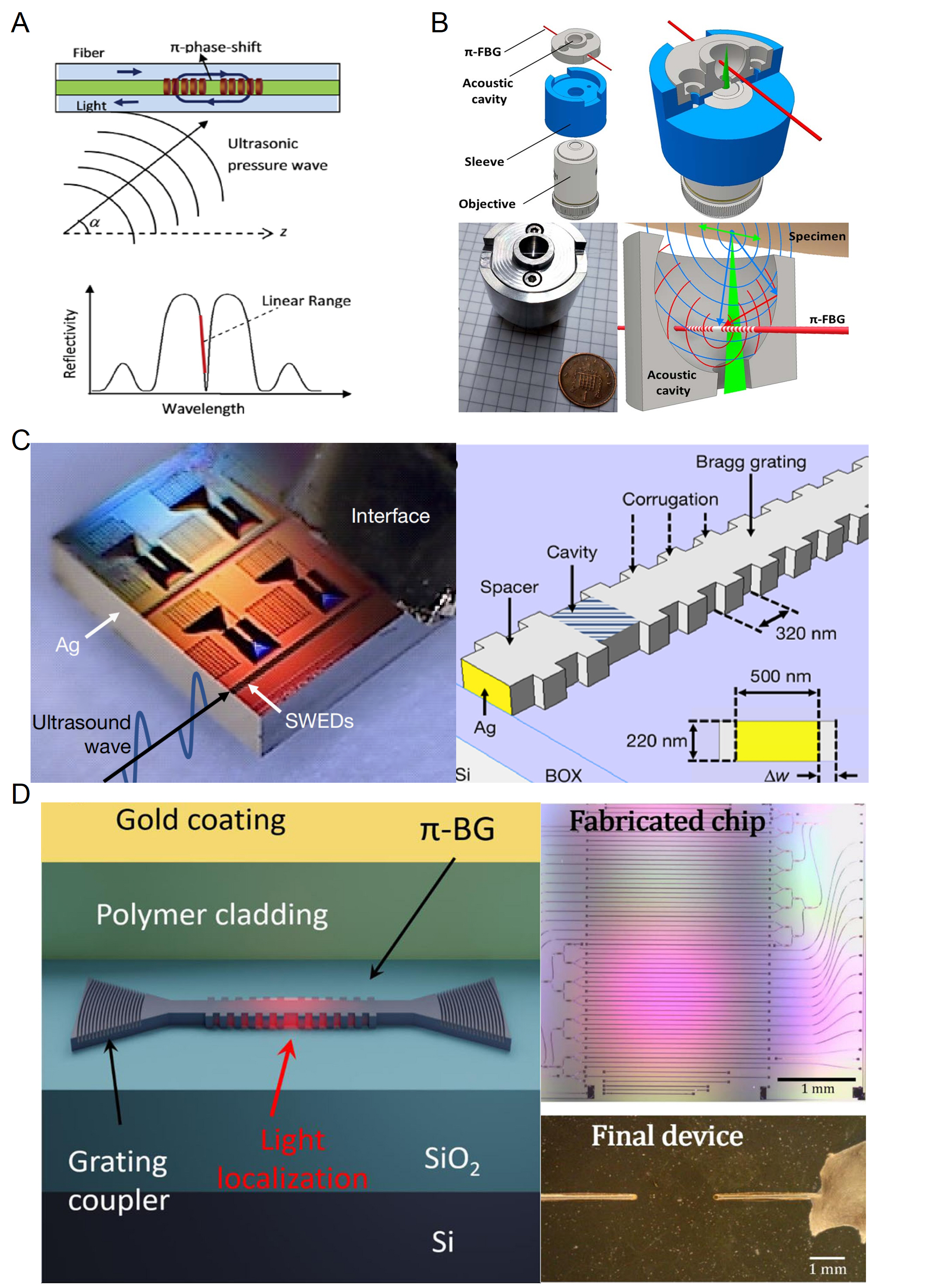}
\end{center}
\caption{$\pi$-phase-shifted Bragg grating ultrasound sensors.
        \textbf{(A)} Schematic of a $\pi$-phase shifted FBG and an ultrasound pressure wave incident onto the grating with an incident angle of $\alpha$ (upper). The lower figure shows the schematic of the reflection spectrum of a $\pi$-FBG.
        \textbf{(B)} Design and operating principle of the $\pi$-FBG-based sensor with an appropriately designed acoustic cavity.
        \textbf{(C)} Design of the silicon waveguide-etalon detector.
        \textbf{(D)} Schematic and the optical images of the silicon Bragg grating ultrasound sensors.
        Reprint \textbf{(A)} from Ref. \citep{FBG3};  
        \textbf{(B)} from Ref. \citep{FBG5}; 
        \textbf{(C)} from Ref. \citep{FBG7};
        \textbf{(D)} from Ref. \citep{FBG8}.}
        \label{fig:7}
\end{figure}

The Bragg grating is a structure with a periodic refractive index. When the Bragg condition is satisfied, there is a high reflectivity in a very small frequency range. The Bragg condition is
\begin{equation}
\lambda_{\text {B}}=2n_{\text {eff }}\Lambda,
\end{equation}
where $\lambda_{\text {B}}$ is the incident light wavelength, $n_\text {eff}$ is the effective refractive index of the Bragg grating and $\Lambda$ is the grating period. When an acoustic wave is applied on a Bragg grating, it changes its effective refractive index and period, leading to a change in the reflectivity of the Bragg grating \citep{FBG10}. However, this approach is the result of interference and does not take advantage of optical resonance. Additionally, only ultrasonic waves with wavelengths longer than the grating length can be accurately detected due to the non-uniformity of ultrasonic waves on the Bragg grating disturbance \citep{FBG9}. So researchers introduce a mutation of the $\pi$ phase in the center of the Bragg grating to make $\pi$-phase-shifted Bragg gratings ($\pi$-BGs). As a consequence of this phase jump, the grating acts as a highly reflective mirror forming an F-P cavity-like structure in the Bragg grating. Figure \ref{fig:7}A shows a schematic diagram of a $\pi$-BG ultrasound sensor and its reflection spectrum \citep{FBG3}. The formation of the resonator introduces a sharp change at the center of the reflection spectrum (denoted in the red line in the reflectivity spectrum in Figure \ref{fig:7}A), which significantly enhances the optical response to ultrasound while reducing the sensing area. In 2011, a $\pi$-phase-shifted fiber Bragg grating ($\pi$-FBG) with a reflectivity of over 90\% was used for ultrasound sensing, which has realized a detection range of 10~MHz and a NEP of 440~Pa \citep{FBG2}. They used a continuous-wave laser to monitor the shift in resonance wavelengths, a method that was influenced by laser noise. To improve the sensitivity of the optical readout, Riob\'{o} et~al. measured the phase change near the resonance using a balanced Mach-Zendel interferometer \citep{FBG6}. The interferometric optical path adjustment allowed the phase noise of the laser to be canceled out, achieving 24 times higher SNR than conventional intensity measurement methods. 

Due to the chip integration capability, there is great potential for $\pi$-BGs to be employed in bio-imaging. In 2016, Wissmeyer et~al. demonstrated all-optical photoacoustic microscopy using a $\pi$-FBG. Photoacoustic images of a mouse ear and a zebrafish larva ex vivo have been attained with the optical resolution \citep{FBG4}. Benefitting from the high optical focusing capability and the wide bandwidth ultrasound inspection capability of the $\pi$-FBG, they achieved a lateral resolution of 2.2~\textmu m and an axial resolution of 10.9~\textmu m. $\pi$-FBGs can also be well combined with optical microscopy to achieve multi-mode imaging. Figure \ref{fig:7}B shows a $\pi$-FBG and an acoustic resonant cavity compactly combined to increase the ultrasound response while allowing convenient integration into any optical microscope. Shnaiderman et~al. used this system to achieve the first in vivo sample measurements in epi-illumination mode with entirely optical hybrid optical and optoacoustic microscopy \citep{FBG5}. 

$\pi$-BG can be implemented not only in optical fibers but also in chip-integrated waveguides, which is $\pi$-phase-shifted waveguide Bragg grating ($\pi$-WBG). Shnaiderman et~al. has taken advantage of the miniaturization of on-chip integration by compressing the sensing area to 200~nm$\times$500~nm and implementing an array of eight sensors \citep{FBG7}. Figure \ref{fig:7}C shows the details of their structures. This sensor has realized a sensitivity of 9~mPa Hz$^{-1/2}$ and a wide bandwidth reaching 230~MHz. Such excellent performance has allowed imaging of features 50 times smaller than the wavelength of ultrasound detected, which has enabled ultrasound imaging at a resolution comparable to that achieved with optical microscopy. A further improvement to the $\pi$-WBG was made by Hazan et~al. in 2022, by coating the grating surface with an elastic medium to eliminate the parasitic effect of surface acoustic waves, as shown in Figure \ref{fig:7}D \citep{FBG8}. This sensor has realized a sensitivity of 2.2~mPa Hz$^{-1/2}$ and a bandwidth of up to 200~MHz. The detector was also demonstrated in vivo for high-resolution optoacoustic tomography using an array of $\pi$-FBGs.

\subsection{Whispering gallery mode microcavities}

WGM was first studied in acoustic waves when Lord Reighley found that he could hear two people whispering even if they were very far away. This is because sound waves can be continuously reflected along the curved wall with very small propagation loss. The concept of WGM was then extended to microwaves and optical waves. Analogous to sound waves, light waves can be confined inside a closed circular high-refractive-index dielectric structure through total internal reflection. When the optical path is equal to an integer number times the optical wavelength, the resonance condition is satisfied. With the development of microfabrication technologies in the past few decades, WGM microcavities with extremely high optical $Q$ factors have been realized \citep{WGM2003,36,CaF,26}. 
With their high optical $Q$ factor, small mode volume, and flexible material systems and shapes, WGM microcavities have been demonstrated in various sensing applications \citep{27,WGMmass,WGMgas,WGMforce,WGMmagnetometry2020,WGMtemperature,44,WGMnanoparticles2022,38,41,42,43,47,WGMparticle3,WGMparticle4,WGMparticle5,WGMparticle2,WGMparticle1,45,39,40,46}. In recent years, WGM microcavities have also been used for ultrasound sensing. In the following, we present the recent ultrasound sensing works using WGM microcavities with different geometries, including microrings, microspeheres, microbubbles, microdisks, and microtoroids. 

\subsubsection{Microrings}

\begin{figure}[h!]
\begin{center}
\includegraphics[width=180mm]{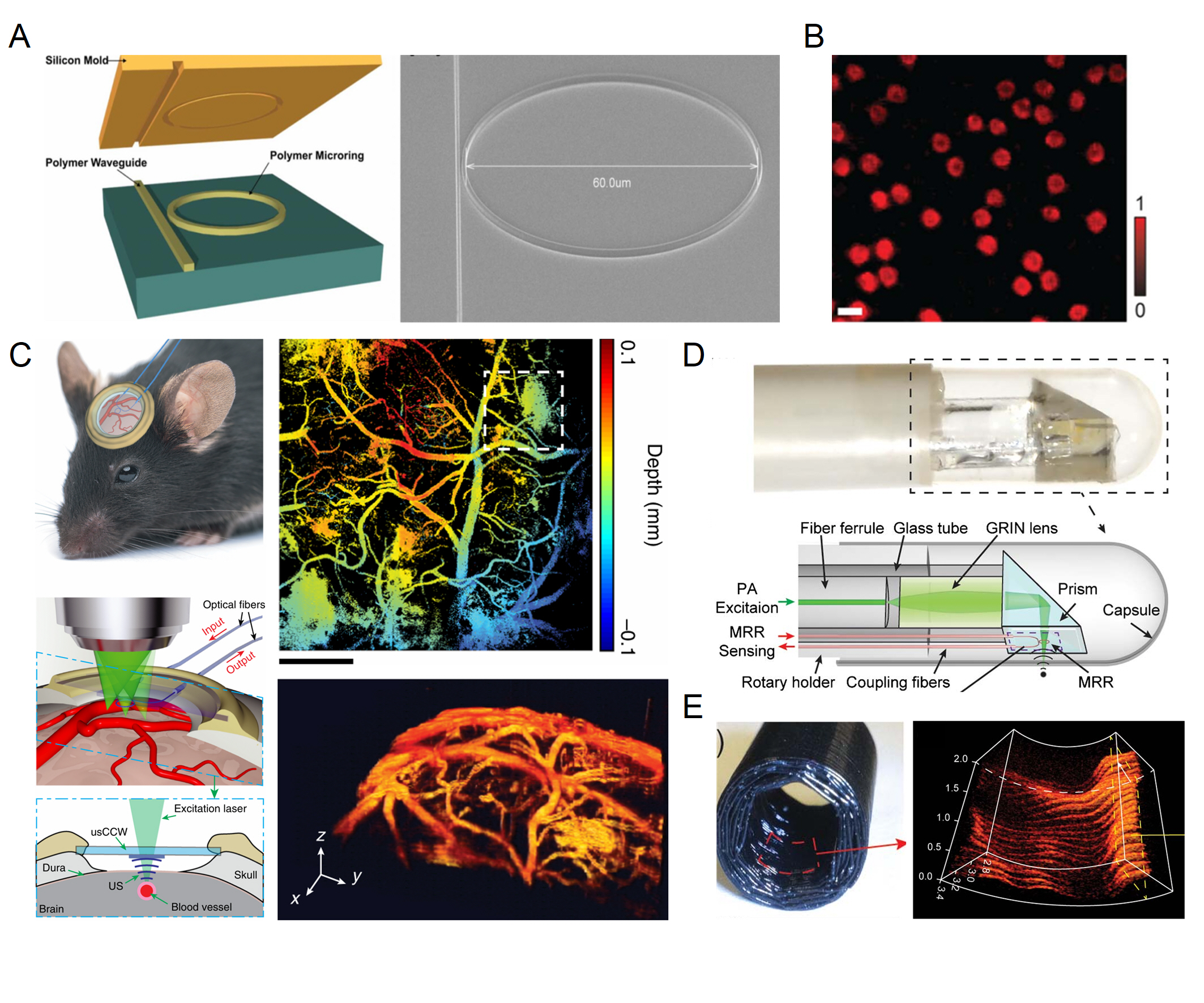}
\end{center}
\caption{Microring ultrasound sensors.
        \textbf{(A)} Schematic of a polymer microring resonator fabricated using nano-imprinting lithography method.
        \textbf{(B)} Photoacoustic microscopy image of single red blood cells in a mouse blood smear.
        \textbf{(C)} In vivo photoacoustic microscopy cortical imaging using an ultrasound-sensing chronic cranial window.
        \textbf{(D)} Photograph and illustration of the microring-based photoacoustic endoscopic probe.
        \textbf{(E)} Photograph of the black plastic tube phantom and its 3D photoacoustic volumetric rendering of its inner surface.
        Reprint \textbf{(A)} from Ref. \citep{25};  
        \textbf{(B)} from Ref. \citep{WGM9}; \textbf{(C)} from Ref. \citep{WGM14}; \textbf{(D-E)} from Ref. \citep{WGM5}.
        }\label{fig:8}
\end{figure}

Microrings are the most used type of WGM microcavities for ultrasound sensing because they can be easily integrated on silicon chips, and have a wide choice of materials. Microrings are generally fixed directly to the substrate and it is difficult for an ultrasound to modulate the cavity length. Polymer materials with low Young's modulus are good choices to increase the strain to improve the response to ultrasound. Some polymer materials such as polymethyl methacrylate (PMMA) \citep{WGM30,PMMA} and SU-8 \citep{WGM8}, can be directly patterned using electron beam lithography (EBL). However, their optical $Q$ factors are usually limited to the $10^3$ to $10^4$ range. Zhang et~al. used a nanoimprinting method that used silicon as a mold (Figure \ref{fig:8}A) to fabricate polystyrene (PS) microrings, and increased the $Q$ factor of polymer microrings to $10^5$ by optimizing the nanoimprinting lithography \citep{25}. Using this high-$Q$ PS microring cavity, they have achieved a broadband response of 350~MHz, with a NEP of 105~Pa in this frequency range. Such a large response bandwidth allowed them to achieve sub-3~\textmu m axial resolution in photoacoustic imaging \citep{WGM6}. They have also explored the potential of an ultrasound sensing array, with a one-dimensional array consisting of four microrings coupled with a single waveguide \citep{WGM2}.

\begin{figure}[h!]
\begin{center}
\includegraphics[width=180mm]{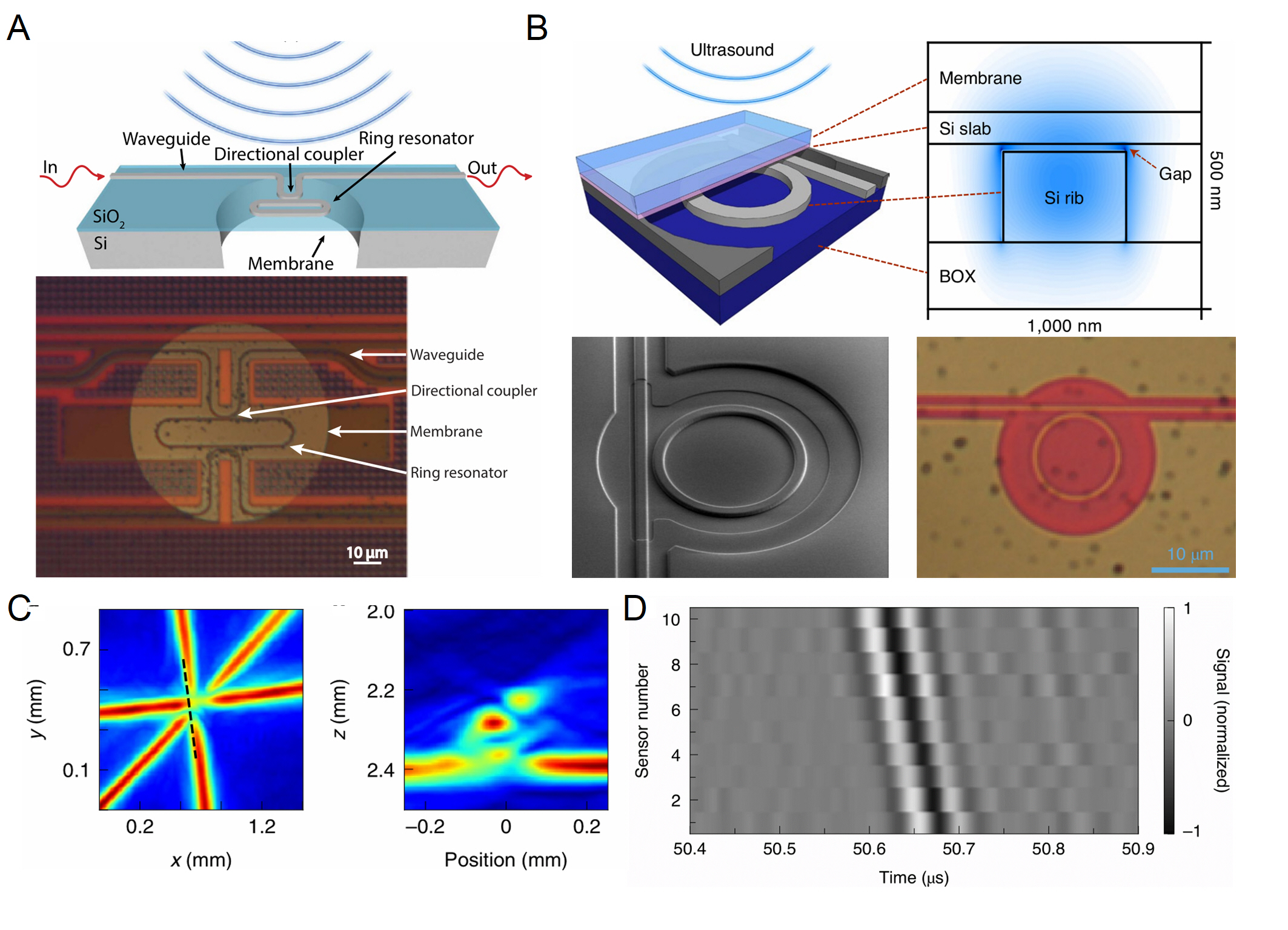}
\end{center}
\caption{
        \textbf{(A)} Schematic and microscopic image of the optical micro-machined ultrasound transducer, showing the photonic microring resonator on top of the membrane.
        \textbf{(B)} Schematic, SEM and optical images of the optomechanical ultrasound sensor, consisting of a thin film coupled with a silicon microring resonator.
        \textbf{(C)} 3D photoacoustic tomographic reconstructions of a phantom consisting of three overlaying polyamide structures.
        \textbf{(D)} Ultrasound signal time traces received by the ten microring ultrasound sensors.
        Reprint \textbf{(A)} from Ref. \citep{WGM11};  
        \textbf{(B-D)} from Ref. \citep{WGM24}.
        }\label{fig:9}
\end{figure}

Polymer microrings have been used in photoacoustic imaging already in 2011, which offer a resolution of 5~\textmu m laterally and 8~\textmu m axially \citep{WGM3}. Using a polymer microring on a microscope coverslip, Li et~al. conceived and fabricated an optically transparent ultrasound detector in 2014 \citep{WGM8}. It enabled high-sensitivity ultrasound detection over a wide receiving angle with a bandwidth of 140~MHz and an estimated NEP of 6.8~Pa. They verified an axial resolution of 5.3~\textmu m by photoacoustic imaging of a carbon-black thin-film target. In 2015, they improved the system to achieve photoacoustic imaging of mouse erythrocytes as shown in Figure \ref{fig:8}B, with an axial resolution of 2.1~\textmu m \citep{WGM9}. In 2019 they reported a disposable ultrasound-sensing chronic cranial window featuring an integrated transparent nanophotonic ultrasound detector \citep{WGM14}. This detector was used to demonstrate photoacoustic microscopy of the cortical vascular network in live mice for 28 days as shown in Figure \ref{fig:8}C. The small size of the microring also allows it to be used as a probe for endoscopy. Dong et~al. have attached optically transparent polymer microrings and prisms together, allowing for a compact structure where excitation and detection are integrated \citep{WGM5}, as shown in Figure \ref{fig:8}D. They then achieved 3D photoacoustic imaging of the inner wall of a black plastic tube as well as the hair by rotating the probe, with the image shown in Figure \ref{fig:8}E.

Silicon microrings have also been widely applied for ultrasound sensing, as the well-developed silicon photonics technology in the past few decades has allowed low-cost and mass fabrication of silicon microring cavities on silicon-on-insulator (SOI) platforms. In order to increase the mechanical compliance and thus the response of the silicon microring to ultrasound, optical micro-machined ultrasound sensors based on acoustic membranes have been developed, by etching away the silicon substrate underneath the silicon microring \citep{WGM11,WGM12}, with the schematic and the optical microscope image of the structure shown in Figure \ref{fig:9}A. Ultrasound pressure down to 0.4~Pa has been realized using this suspended silicon membrane \citep{WGM11}. The coupling region may also be deformed in this way resulting in a nonlinear readout. Then Yang et~al. proposed to partially etch out the silicon substrate under the microring region to maintain a linear readout \citep{WGM19}. In 2021, Westerveld et~al. fabricated a thin silicon film over a silicon microring with a 15~nm air gap between them, where ultrasound can change the air gap by causing the thin film to vibrate thus affecting the intracavity optical field of the microring \citep{WGM24}. An illustration of this structure, as well as its SEM and optical microscope images, are shown in Figure \ref{fig:9}B. Using a microring with a diameter of 20~\textmu m, NEP of 1.3~mPa Hz$^{-1/2}$ has been realized in the 3-30~MHz frequency range, which is dominated by acoustomechanical noise. Figure \ref{fig:9}C shows the results of their 3D photoacoustic imaging of a phantom consisting of three overlaying polyamide structures, obtained using this ultrasound sensor. They have also designed a one-dimensional array of ten microrings with their resonance wavelengths uniformly distributed over a free spectral range of 17~nm. The feasibility of the array detection was verified by measuring the delay in the response of different microrings to the ultrasound, as shown in the ultrasound signal time traces in Figure \ref{fig:9}D. There are also microring-like structures such as microknots that can be used for ultrasound sensing \citep{WGM10,WGM17}.

\subsubsection{Microspheres}

\begin{figure}[h!]
\begin{center}
\includegraphics[width=180mm]{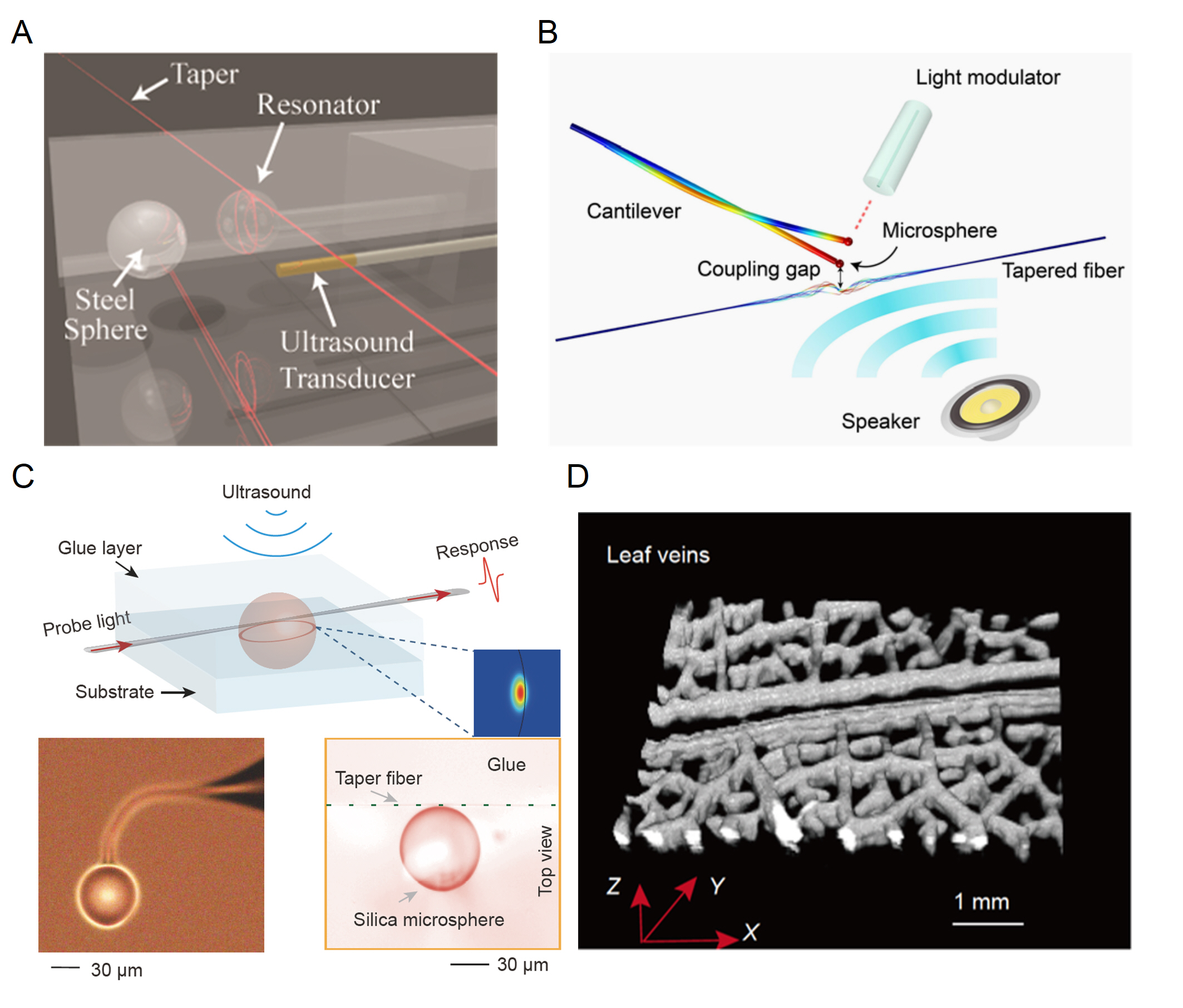}
\end{center}
\caption{Microsphere ultrasound sensors.
        \textbf{(A)} Rendering of the microsphere ultrasound sensor setup.
        \textbf{(B)} Schematic illustration of the mechanical modes of the cantilever-microsphere coupled structure, excited by a temporally-modulated laser beam and a sound wave.
        \textbf{(C)} Schematic of microsphere for ultrasound detection and the optical microscopic images of a silica microsphere.
        \textbf{(D)} 3D photoacoustic imaging result of leaf veins.  
        Reprint \textbf{(A)} from Ref. \citep{WGM7};  
        \textbf{(B)} from Ref. \citep{WGM18}; 
        \textbf{(C-D)} from Ref. \citep{WGM26}.
        }\label{fig:10}
\end{figure}

The microsphere is another commonly used device geometry as it can be easily fabricated. For instance, silica microspheres can be fabricated by melting the end of a fiber tip using a CO$_2$ laser or a fusion splicer. Microspheres are also often used for acoustic wave measurements \citep{WGM30,AAPPS}. In 2014, Chistiakova et~al. performed ultrasound sensing in water using an ultra-high $Q$ silica microsphere \citep{WGM7}. They have simulated and experimentally verified that the microspheres can detect the echo signals of steel balls as well as water tanks as shown in Figure \ref{fig:10}A. In 2020, Yang et~al demonstrated an optomechanical microdevice based on Brillouin lasing in an optical microcavity as a sensitive unit to sense external light, sound, and microwave signals in the same platform \citep{WGM18}, with the structure shown in Figure \ref{fig:10}B. They have achieved a NEP of 267~\textmu Pa Hz$^{-1/2}$, corresponding to a minimum detectable force of 10~pN Hz$^{-1/2}$. The suspended microsphere structure utilized the mechanical vibration modes of the fiber to improve sensitivity. A thin fiber taper is usually used to couple light into the microcavity, and the coupling strength is very sensitive to the distance between the fiber taper and the microcavity. The displacement of the fiber taper caused by ultrasound is much greater than that of the microsphere, so measuring the change in the spacing between the fiber taper and the microsphere is an effective detection mechanism. This dissipative coupling mechanism was then investigated using microspheres by Meng et~al. \citep{WGM27}. They found that the response to ultrasound through dissipative coupling was two orders of magnitude higher than the dispersive coupling mechanism. In order to make the microsphere ultrasound sensor more compact and robust to environmental disturbance, Sun et~al. used glue to encapsulate the microspheres and fibers to avoid contamination \citep{WGM26}, with the schematic illustration as well as the optical microscope images of the sensor before and after the encapsulation are shown in Figure \ref{fig:10}C. A NEP of as low as 160~Pa at 20~MHz was achieved with the ultrasound response up to 70~MHz. As shown in Figure \ref{fig:10}D, they have also used this sensor to successfully perform a 3D photoacoustic imaging of leaf veins. Ultrasound sensing in underwater environments has also been demonstrated using packaged microspheres \citep{WGM25}.

\clearpage

\subsubsection{Microbubbles}

\begin{figure}[h!]
\begin{center}
\includegraphics[width=180mm]{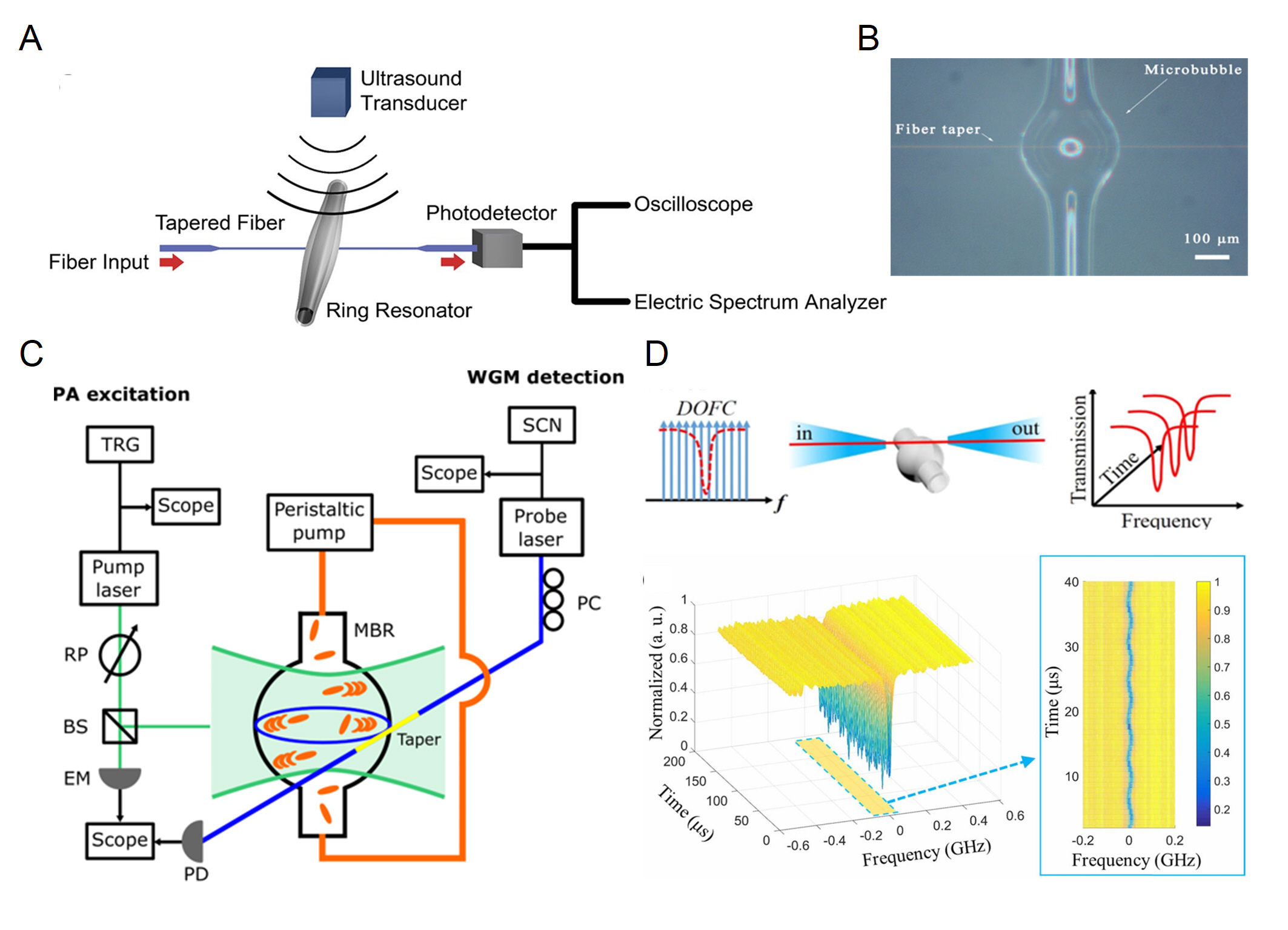}
\end{center}
\caption{Microbubble ultrasound sensors.
        \textbf{(A)} Schematic of the experimental setup for ultrasound detection using a microbubble.
        \textbf{(B)} Optical micrograph of optical microbubble sensor for underwater acoustic detection.
        \textbf{(C)} Schematic of the experimental setup to detect the photoacoustic signal generated by plasmonic nanoparticles.
        \textbf{(D)} Schematic of the experiments based on digital optical frequency comb methods (upper). The lower figure shows the intensity responses to ultrasonic waves.
        Reprint \textbf{(A)} from Ref. \citep{WGM13};  
        \textbf{(B)} from Ref. \citep{WGM28}; 
        \textbf{(C)} from Ref. \citep{WGM22}; 
        \textbf{(D)} from Ref. \citep{WGM21}.
        }\label{fig:11}
\end{figure}

Both microrings and microspheres are solid microcavities that are less prone to deform than hollow structures. As a result, microbubble cavities fabricated using hollow capillaries have been extensively used for ultrasound sensing, and the walls of the capillaries can be made very thin, greatly enhancing the response to ultrasound. In 2017, Kim et~al. developed a microbubble-based ultrasound sensor (Figure \ref{fig:11}A) that has reached a NEP of 215~mPa Hz$^{-1/2}$ and 41~mPa Hz$^{-1/2}$ at 50~kHz and 800~kHz in air, respectively \citep{WGM13}. Microbubbles also use fiber tapers to couple light in and out, and need to be encapsulated in complex detection environments. Tu et~al. used the encapsulated microbubble (Figure \ref{fig:11}B) to detect acoustic waves at low frequencies in the 10~Hz to 100~kHz range and achieved a NEP of 2.2~mPa Hz$^{-1/2}$ \citep{WGM28}. Benefiting from their encapsulated structure, microbubble sensors exhibited stable performance at different temperatures and static pressures. One advantage of microbubbles that distinguishes them from other microcavities is that their walls can serve as ultrasound transducers and the hollow structure inside acts as a sample container. In recent years, there have also been numerous works \citep{WGM16,WGM22,WGM23} that achieve photoacoustic detection of flowing samples by injecting nanoparticles into microbubbles, as shown in Figure \ref{fig:11}C. This approach not only allows non-contact detection of the target particles but also allows the optical absorption spectra to be distinguished between different particles. In 2020, Pan et~al. used a microbubble cavity combined with a digital optical frequency comb for ultrasound detection in air, which allows the full mode spectrum to be obtained at a timescale of the \textmu s level. The working principle and experimental results of this work are shown in Figure \ref{fig:11}D \citep{WGM21}. They have achieved a NEP of 4.4~mPa Hz$^{-1/2}$ in the air at 165~kHz frequency. They also achieved a high positioning precision by measuring the phase difference between two microbubbles. The optical frequency comb was also used in microrings on the chip for ultrasound measurement \citep{WGM20}.

\subsubsection{Microdisks and microtoroids}

\begin{figure}[h!]
\begin{center}
\includegraphics[width=180mm]{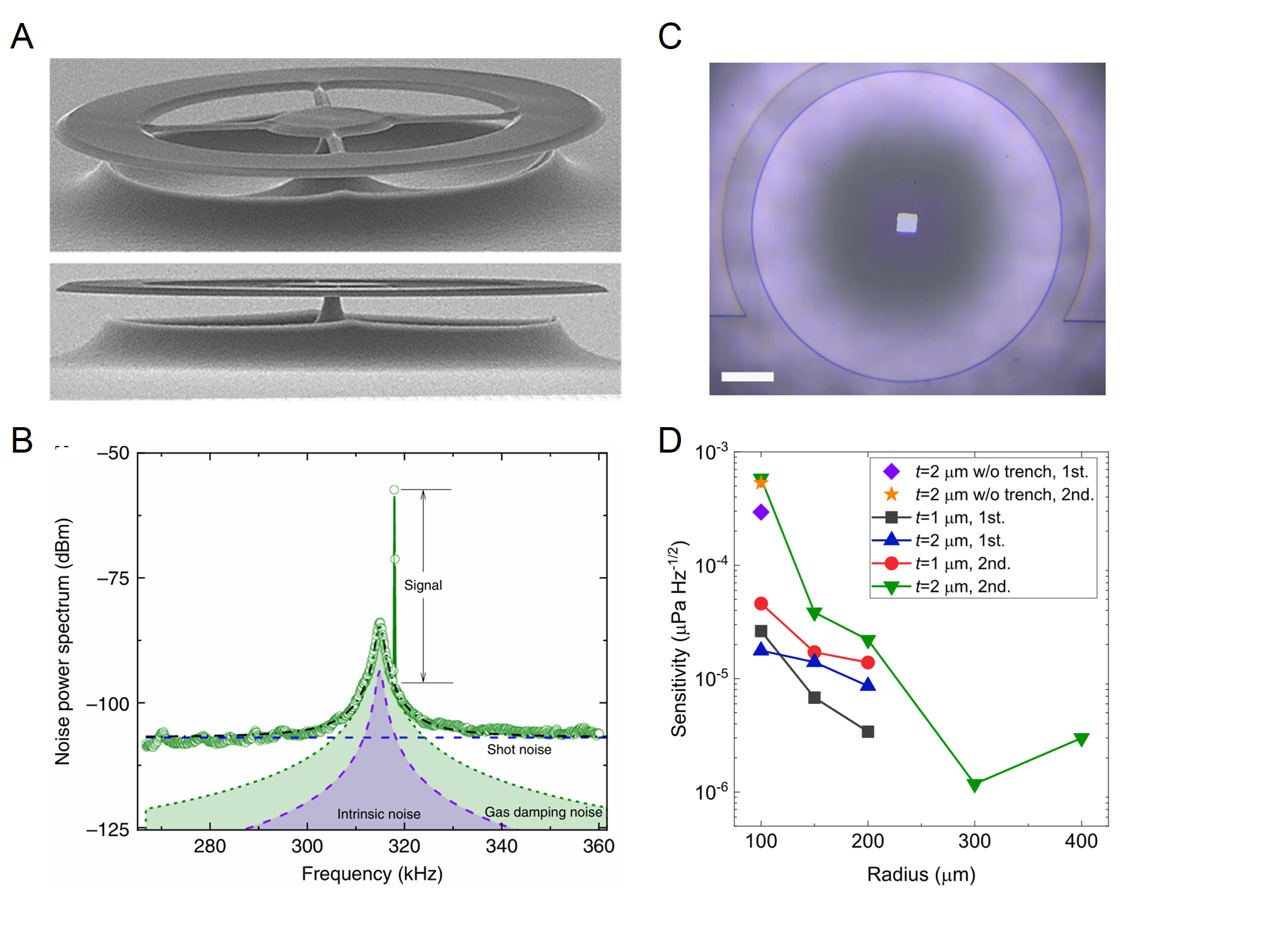}
\end{center}
\caption{Microdisk ultrasound sensors.
        \textbf{(A)} SEM picture of a suspended spoked microdisk.
        \textbf{(B)} Noise power spectrum of the sensor near a mechanical mode of the microdisk.
        \textbf{(C)} Top-view optical microscope image of a microdisk with a trench structure.
        \textbf{(D)} Sensitivities at the flapping modes of microdisks with different thicknesses and radii.
        Reprint \textbf{(A-B)} from Ref. \citep{WGM15};  
        \textbf{(C-D)} from Ref. \citep{28}.
        }\label{fig:12}
\end{figure}

\begin{figure}[h!]
\begin{center}
\includegraphics[width=85mm]{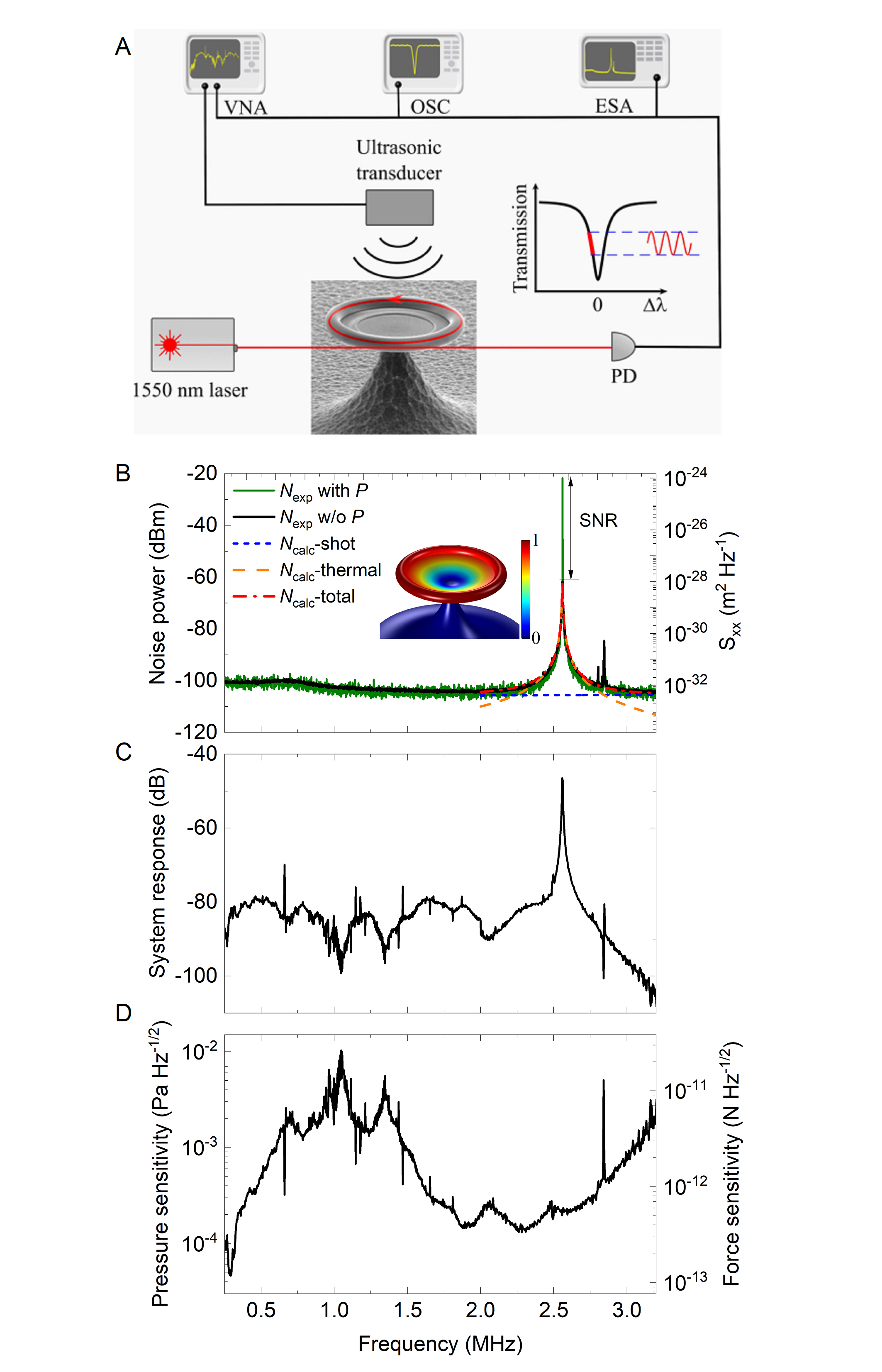}
\end{center}
\caption{Microtoroid ultrasound sensors.
        \textbf{(A)} Schematic diagram of the experimental setup for ultrasound sensing using a microtoroid.
        \textbf{(B,C,D)} Noise power spectra, system response, and sensitivity spectra of the microtoroid ultrasound sensor.
        Reprint \textbf{(A-D)} from Ref. \citep{WGM29}.
        }\label{fig:13}
\end{figure}

In addition to the above-mentioned three common types of WGM microcavities, microdisks have also been utilized for ultrasound sensing. There are several advantages of the microdisk structure. First, with the current microfabrication techniques, the sensing areas of microdisks can be easily made very large, which is helpful for improving the sensitivity. Second, the suspended microdisk structures increase their mechanical compliance thus enhancing their response to ultrasound, and decrease their mechanical damping rate $\gamma$ thus allowing better thermal-noise-limited sensitivity (Eq. (4)). In 2019, Basiri-Esfahani et~al. demonstrated an ultrasound sensor using a suspended spoked microdisk, and reached the noise region dominated by collisions of gas molecules \citep{WGM15}. SEM images of the suspended spoked microdisk are shown in Figure \ref{fig:12}A. The spoke structure can make the microdisk more mechanically compliant, reducing mechanical losses and making it easier to reach the thermal-noise-limited regime. Figure \ref{fig:12}B shows the noise power spectrum (black) of the microdisk around the mechanical mode as well as its ultrasound response at a single frequency (green curve). This allowed NEPs of 8-300~\textmu Pa Hz$^{-1/2}$ at a range of frequencies between 1~kHz and 1~MHz. They used both dissipative and dispersive mechanisms to read out the different mechanical vibration modes. This study demonstrated a significant improvement in the sensitivity of ultrasound sensors in the range dominated by thermal noise. In 2022, Yang et~al. performed a more systematic study on the thermal-noise-limited ultrasound sensitivity using suspended microdisks both theoretically and experimentally \citep{28}. The sensitivity was optimized by varying the radius and thickness of the microdisk as well as using a trench structure around the disk. Top-view optical microscope image of a microdisk with a trench structure is shown in Figure \ref{fig:12}C. Sensitivities of microdisks with different thicknesses and radii are shown in Figure \ref{fig:12}D. A peak sensitivity of 1.18~\textmu Pa Hz$^{-1/2}$ has been realized at 82.6~kHz, using a microdisk with a radius of 300~\textmu m and a thickness of 2~\textmu m. 

The optical $Q$ factors can be further improved by melting the edge of the microdisk to form a microtoroid with an atomically smooth surface and $Q$ factors as high as 10$^8$ \citep{WGM2003}. To further expand the frequency range of air-coupled ultrasound detection, Yang et~al. then used microtoroids to improve the megahertz-frequency ultrasound detection \citep{WGM29}, with the ultrasound measurement setup shown in Figure \ref{fig:13}A. Using a microtoroid with a very thin silicon pedestal, they have achieved a mechanical $Q$ factor of 700 at the first-order flapping mode at 2.56~MHz. Figure \ref{fig:13}B shows the noise power spectra of the microtoroid ultrasound sensor with (green curve) and without (black curve) applying ultrasound at 2.56~MHz, with the displacement PSD ($S_{xx}$) shown on the right axis. Figures \ref{fig:13}C shows the ultrasound response at different frequencies, and Figure \ref{fig:13}D shows the pressure and force sensitivities on the left and right axes respectively. Thermal-noise-limited sensitivity has been achieved in a frequency range of 0.6~MHz near the mechanical resonance. Sensitivities of 46~\textmu Pa Hz$^{-1/2}$-10~mPa Hz$^{-1/2}$ have been realized in the frequency range of 0.25-3.2~MHz. 

\subsection{Performance comparison}

\renewcommand\arraystretch{1.5}
\begin{table}[htbp]
\caption{Summary of the performances of optical ultrasound detection with different microcavities}
\tabcolsep=4mm
\begin{tabular}{ccccccc}

\hline
      & Structure                                                                 & $Q$ factor         & \begin{tabular}[c]{@{}l@{}}CF \\ (MHz)\end{tabular} & \begin{tabular}[c]{@{}l@{}}BW at\\ -6 dB (MHz)\end{tabular} & \begin{tabular}[c]{@{}l@{}}NEPD\\ (mPa Hz$^{-1/2}$)\end{tabular} & \begin{tabular}[c]{@{}l@{}}Acceptance\\ angle\end{tabular} \\ \hline
FP    &  Diaphragm \citep{FP2}         & -                & 0.01                                                          & 0.022                                                        & 0.060                                                        & Around 80$^{\circ}$                                                 \\
      & F-P etalon \citep{FP4}                                                       & -                & 1                                                             & 22.5                                                         & 0.450                                                        &  \textless 20$^{\circ}$                                                \\
      & \begin{tabular}[c]{@{}l@{}}Plano-concave \\ microresonator \citep{FP6}\end{tabular}   & \textgreater 10$^5$ & 3.5                                                           & \textgreater 20                                                          & 2.1                                                          & 180°                                               \\
      & \begin{tabular}[c]{@{}l@{}}Microbubble \\ F-P cavity \citep{FP13}\end{tabular} & -                & 0.7                                                           & 0.8                                                          & 3.4                                                          & 180$^{\circ}$                                                 \\
      & \begin{tabular}[c]{@{}l@{}}Buckled-dome \\ microcavities* \citep{FP16}\end{tabular}    & $\sim$10$^3$        & -                                                             & \textgreater15                                                          & 0.03-0.1                                                     & 120$^{\circ}$                                     \\         \hline
$\pi$-BG & FBG \citep{FBG2}                                                                      & 2$\times$10$^5$            & 6.5                                                           & 3                                                            & (NEP) 440~Pa                                                  & -                                                          \\
      & FBG \citep{FBG5}                                                                       & 1.9$\times$10$^5$        & $\sim$25                                                      & $\sim$36                                                     & (NEP) 88~Pa                                                    & 153$^{\circ}$                                                       \\
      & WBG \citep{FBG7}                                                                      & $\sim$10$^5$        & -                                                             & 230                                                          & 9                                                            & 148$^{\circ}$                                                       \\
      & WBG \citep{FBG8}                                                                      & -                & -                                                             & 200                                                          & 2.2                                                          & -                                                          \\ \hline
WGM   & Microring \citep{WGM6}                                                              & $\sim$10$^5$        & -                                                             & 350 (-3~dB)                                                    & (NEP) 105~Pa                                                  & -                                                          \\
      & Microring \citep{WGM9}            & $\sim$10$^4$        & -                                                             & 280                                                          & (NEP) 6.8~Pa                                                   & 14$^{\circ}$                                               \\
      & Microring \citep{WGM11}              & $\sim$10$^4$        & 0.76                                                          & 0.14                                                         & (NEP) 0.4~Pa                                                   & -                                                          \\
      & Microring \citep{WGM24}             & $\sim$10$^4$        & -                                                             & 27                                                           & 1.3                                                          & 120°                                               \\
      & Microsphere \citep{WGM7}                                                               & 9.5$\times$10$^7$          & 40                                                            & 5                                                            & (NEP) 0.535~Pa                                                & -                                                          \\
      & Microsphere \citep{WGM18}                                                               & 10$^8$              & 0.0057                                                        & -                                                            & 0.267                                                        & -                                                          \\
      & Microsphere \citep{WGM26}                                                           & $\sim$10$^5$        & 20                                                            & 70                                                           & (NEP) 100~Pa                                                   & -                                                          \\
      & Microsphere \citep{WGM27}                                                              & $\sim$10$^6$        & 0.14                                                          & -                                                            & 1.29                                                         & -                                                          \\
      & Microbubble \citep{WGM13}                                                               & 3.5$\times$10$^7$         & 0.8                                                           & 0.2                                                          & 41                                                           & -                                                          \\
      & Microbubble \citep{WGM21}                                                               & 3$\times$10$^7$          & 0.165                                                         & -                                                            & 4.4                                                          & -                                                          \\
      & Microbubble \citep{WGM28}                                                               & 5.2$\times$10$^5$         & 0.001                                                         & 0.1                                                 & 2.2                                                          & 105.5$^{\circ}$                                                     \\
      & Microdisk \citep{WGM15}                                                                & 3.6$\times$10$^6$         & 0.318                                                         & -                                                            & 0.008-0.3                                                    & -                                                          \\
    & Microdisk \citep{28}                                                                & 3$\times$10$^6$         & 0.0826                                                         & -                                                            & 0.00118                                                   & -                                                          \\
      & Microtorid \citep{WGM29}                                                                & $\sim$10$^7$        & 2.56                                                          & 1.13                                                          & 0.046-10                                                     & -                                                          \\ \hline

\end{tabular}
\begin{tablenotes}
\footnotesize
\item CF: center frequency; BW: bandwidth
\end{tablenotes}
\end{table}\label{table:1}

Table 1 summarizes the main parameters of the three types of optical microcavity-based ultrasound sensors, including optical $Q$ factor, center frequency, bandwidth, NEPD or NEP, and acceptance angle. F-P cavities can achieve an optical $Q$ factor of about 10$^4$ to 10$^5$ with a highly reflective dielectric layer. The optical $Q$ factor is at the same level for $\pi$-BGs, and their $Q$ factors can be further increased by increasing the grating length. Among the three types of optical microcavites, WGM microcavities can achieve the highest optical $Q$ factors, especially for microspheres, microbubbles, and microtoroids, which are at the level of $10^7$-$10^8$. Selecting materials with lower optical absorption losses, such as silicon nitride, can improve the optical $Q$ factors of microrings \citep{37}. Higher optical $Q$ factors allow the thermal-noise-limited sensitivity to be achieved, leading to a lower NEP. Most of the sensors with NEP to the micropascal level have exploited the mechanical resonances of the structures to achieve the thermal-noise-limited sensitivity, except for a few millimeter-scale F-P cavities. However, mechanical resonances can lead to smaller detection bandwidth, which can affect certain applications. For most imaging applications, bandwidth at the megahertz level is required, but only some sensors can detect megahertz frequency ultrasound in water, due to the large propagation loss of high-frequency ultrasound waves. Air-coupled high-sensitivity ultrasound detection above 1~MHz frequency was only realized using a microtoroid cavity because of the even larger absoprtion loss in air. Transducers on optical fibers generally have wider acceptance angles and are better suited to receive ultrasound signals from multiple directions. The fiber itself is a good optical transmission device and is easy to connect to external devices, such as lasers. Sensors integrated onto a chip have slightly narrower acceptance angles but offer advantages of low cost, low power consumption, and mass production. On the other hand, stand-alone sensors face the challenge of stabilizing packaging for practical applications outside of the laboratory. For 2D and 3D imaging, the use of multiple sensor arrays working simultaneously can reduce the need for mechanical moving parts and shorten the imaging time. While 2D array multi-channel parallel sensing has already been achieved with F-P cavities on optical fibers, the development of array sensing is still in its infancy, as only one-dimensional arrays of $\pi$-BGs and WGM microcavities have been demonstrated so far.

\section{Conclusion}

Over the past few decades, optical ultrasound sensors have merged and demonstrated superior sensitivity and bandwidth compared to traditional piezoelectric sensors. Among them, recently developed optical microcavity ultrasound sensors offer the advantages of high sensitivity, broad bandwidth, and miniaturization, and have extensive applications in ultrasound imaging, photoacoustic sensing, etc. This review discusses ultrasound sensing work using optical microcavities. At the beginning, we introduce the sensing principles and the readout mechanisms and discuss the key parameters of microcavity ultrasound sensors. Previous work has shown that thermal noise is the fundamental limitation of NEP, and in this review, we have discussed the factors that influence the sensor response and sensitivity. We then present a comprehensive overview of the works on ultrasound sensing using three different types of optical microcavities, including F-P cavities, $\pi$-phase-shifted Bragg gratings, and WGM microcavities. A performance comparison of different microcavity ultrasound sensors is also provided. While F-P cavity-based ultrasound sensors have low NEPs, they usually require suspended thin film structures and have a relatively large sensing area. In contrast, solid F-P cavities have inferior sensitivities but much larger response bandwidth. Fiber-based F-P cavities allow almost full spatial angle response and multi-channel parallel sensing. $\pi$-BGs have the advantages of broadband response, large acceptance angle, multi-parameter sensing, and ease of on-chip integration, but require improved sensitivity. WGM microcavities can achieve higher sensitivities, due to the higher optical and mechanical $Q$ factors which allow thermal-noise-limited sensitivities to be reached. However, WGM microcavities are more sensitive to the direction of ultrasonic waves, making it difficult to achieve a full-angle response. With the development of both science and technology, the performance of optical microcavity ultrasound sensors is expected to further improve, with lower NEPs, broader bandwidths, and larger acceptance angles. Their potential for parallel sensing needs to be further exploited, for high-speed imaging and sensing applications. This could be realized by combining multi-wavelength frequency comb sources \citep{combsensing} with an ultrasound sensor array. In the near future, We expect that optical microcavities will bring new opportunities for ultrasound sensing in applications such as photoacoustic imaging and non-destructive detection, etc.

\section*{Conflict of Interest}

The authors declare that the research was conducted in the absence of any commercial or financial relationships that could be construed as a potential conflict of interest.

\section*{Author Contributions}
XC conducted the literature research, prepared the figures and tables, and wrote the manuscript. HY assisted with the manuscript writing and figure preparation. BBL supervised the manuscript writing process, including the structure design and revision, and provided critical feedback. All authors participated in the manuscript revision, reviewed the final draft, and gave their approvals for submission.

\section*{Funding}
This work is supported by The National Key Research and Development Program of China (2021YFA1400700), the National Natural Science Foundation of China (NSFC) (62222515, 12174438, 91950118, 11934019), and the basic frontier science research program of Chinese Academy of Sciences (ZDBS-LY- JSC003). This work is also supported by the Micro/nano Fabrication Laboratory of Synergetic Extreme Condition User Facility (SECUF).

\bibliography{References}

\end{document}